\documentclass[aps,pra,twocolumn,nofootinbib]{revtex4-2}
\usepackage[colorlinks]{hyperref}
\usepackage[utf8]{inputenc}
\usepackage{graphicx}
\usepackage{multirow}
\usepackage{amssymb}
\usepackage{amsmath}
\usepackage{microtype}

\begin{document}
	
\title{Nuclear polarization effects in atoms and ions}
	
\author{V.~V.~Flambaum$^{1,2,3}$}
\author{I.~B.~Samsonov$^1$}
\author{H.~B.~Tran Tan$^1$}
\author{A.~V.~Viatkina$^{2,3}$}
\affiliation{$^1$School of Physics, University of New South Wales, Sydney 2052, Australia}
\affiliation{$^2$Helmholtz Institute Mainz, GSI Helmholtzzentrum für Schwerionenforschung, 55099 Mainz, Germany}
\affiliation{$^3$Johannes Gutenberg University Mainz, 55099 Mainz, Germany}
	
\begin{abstract}
In heavy atoms and ions, nuclear structure effects are significantly enhanced due to the overlap of the electron wave functions with the nucleus. This overlap rapidly increases with the nuclear charge $Z$. We study the energy level shifts induced by the electric dipole and electric quadrupole nuclear polarization effects in atoms and ions with $Z \geq 20$. The electric dipole polarization effect is enhanced by the nuclear giant dipole resonance. The electric quadrupole polarization effect is enhanced because the electrons in a heavy atom or ion move faster than the rotation of the deformed  nucleus, thus experiencing significant corrections to the conventional approximation in which they `see' an averaged nuclear charge density. The electric nuclear polarization effects are  computed numerically for $1s$, $2s$, $2p_{1/2}$ and high $ns$ electrons. The results are fitted with elementary functions of nuclear parameters (nuclear charge, mass number, nuclear radius and deformation). We construct an effective potential which models the energy level shifts due to nuclear polarization. This effective potential, when added to the nuclear Coulomb interaction, may be used to find energy level shifts in multi-electron ions, atoms and molecules. The fitting functions and effective potentials of the nuclear polarization effects are important for the studies of isotope shifts and nonlinearity in the King plot which are now used to search for new interactions and particles. 
\end{abstract}
	
\maketitle
	
\section{Introduction}
	
Hydrogen-like ions represent a powerful tool for studying various aspects of quantum electrodynamics (QED) and physics beyond the Standard Model (SM). Since these systems are, at least at the electronic level, free from many-body interactions, the spectra of hydrogen-like ions may be calculated with high accuracy, see, e.g., Refs.~\cite{Mohr-review,Shabaev-review} for a review. Corrections to the electronic energy levels in these ions including nuclear recoil, nuclear finite size corrections, one- and two-loop QED corrections (Lamb shift) and nuclear polarization effects have been identified by the ever increasing accuracy of modern experimental techniques, see, e.g., Refs.~\cite{doi:10.1063/1.3120005,KLUGE200883,PhysRevLett.94.223001}.
In this paper, we study the effects of nuclear polarization on the energy spectra of  hydrogen-like ions with $Z \geq 20$. We also seek to extend the formalism for hydrogen-like ions to the case of multielectron atoms.
	
The effects of nuclear polarization are significantly enhanced in a heavy ion because its $s$ and $p_{1/2}$ electron wave functions have sizable overlaps with the nucleus. The technique for computing corrections to atomic spectra due to the electric nuclear polarization was developed in a series of papers \cite{plunien_nuclear_1989,plunien_nuclear_1991,plunien_nuclear-polarization_1995,plunien_erratum:_1996,nefiodov_nuclear_1996}. Therein, it was demonstrated that the leading corrections to an energy level arise from mixing with the nuclear electric giant dipole resonance state (E1) and mixing with nearby nuclear rotational states (E2). The latter mechanism may play a significant role in deformed nuclei where the effect is enhanced by close nuclear rotational levels: in very heavy atoms, these intervals become smaller than typical energy intervals for virtual electron excitations. The goal of this paper is to find the corrections to the atomic energy levels due to these mechanisms for all medium and heavy atoms. The results are presented in terms of simple interpolation formulas which depend on the nuclear charge $Z$, nucleon number $A$,  nuclear radius $R$ and quadrupole deformation parameter $\beta_2$.
	
An important feature of a nuclear giant dipole resonance is that its energy and transition strength are, to a good approximation, monotonic functions of the atomic number $Z$ and mass number $A$. As a result, the energy levels shifts caused by virtual nuclear giant dipole resonance transitions should also be well-described by functions which are monotonic in these parameters. In this paper, we numerically calculate these shifts for a variety of ions with ($Z\geq20$) and fit the results with elementary functions of $Z$, $A$ and $R$. These interpolating functions describe the corresponding energy shifts in all heavy nuclei with a good accuracy: the error is under 2\%, as compared with the results of direct numerical calculations. 
	
In a similar way, we fit the results of numerical calculations for the energy shifts due to nuclear rotational E2 polarization. The error of the interpolating functions in this case is also under 2\%. Note that because the transition strengths of the nuclear rotational E2 transitions have strong dependence on the nuclear deformation parameter $\beta_2$, which changes significantly even between neighboring nuclei, the energy shifts are  non-monotonic functions of $Z$ and $A$. This behaviour of the energy shifts due to the nuclear polarization is expected to give significant contributions to the nonlinearity of the King plot for isotope shifts \cite{flambaum_isotope_2018}.
	
In multielectron atoms and molecules, it is convenient to describe the corrections to the spectra in terms of effective interactions $V_L(r)$ which should be added to the nuclear Coulomb potential. By definition, the expectation values of the potentials $\langle V_L(r)\rangle$ are equal to the energy shifts due to the dipole and quadrupole nuclear polarizations. These potentials may be useful in the study of the nonlinearity of the King's plot \cite{king1963,king_isotope_2013,Gebert2015,Knollmann2019,Manovitz2019}, which  provides information about physics beyond the SM \cite{berengut2018,Frugiuele2017,flambaum_isotope_2018,Yerokhin2020,berengut_generalized_2020,berengut2020}. For example, the  nonlinearity may be interpreted as a manifestation of a new boson mediating electron-nucleus interaction. The study of such non-SM nonlinear deviations would strongly benefit from the careful accounting of SM contributions to the isotope shift nonlinearity. The current paper provides important estimate for the contributions due to the nuclear structure effects in atomic spectra which would be subtracted from experimental data to identify the strength of new, non-SM interactions. 
	
The rest of the paper is organized as follows. In the next section, we review the scalar and tensor nuclear polarizabilities and estimate the effect of the latter in the spectra of heavy atoms. In Sec.~\ref{sec:heavy}, we calculate specific energy shifts in the spectra of medium and heavy hydrogen-like ions and multielectron atoms due to the scalar nuclear polarization. The results for ions are represented in terms of interpolating formulas which reproduce these energy shifts with good accuracy. The effective potentials that produce the energy shifts in multieletron atoms are the subject of Sec.~\ref{Eff_Pot_Sec}. In Sec.~\ref{Conclusion}, we summarize and discuss our findings.
	
In this paper, we use natural units wherein $\hbar=c=1$. Nuclear energies are denoted by the Latin letter $E$ whereas for atomic energy levels we use the Greek letter $\varepsilon$.
	
\section{Tensor nuclear polarizability contribution to the hyperfine  splitting of atomic energy levels}
	
In this section, we compare the effects of the hyperfine splitting of atomic energy levels due to nuclear tensor polarizability with those due to nuclear electric quadrupole  moment. We start with a short review of the nuclear quadrupole moment and its contribution to the atomic hyperfine structure. Then we consider similar contributions from the nuclear tensor polarizability and estimate the ratio between  parameters of these two effects.
	
\subsection{Hyperfine splitting due to nuclear quadrupole deformation}\label{Hyperfine splitting due to nuclear quadrupole deformation}
	
In this subsection, we review the well-known results concerning the contributions of the nuclear quadrupole moment to the hyperfine level splitting \cite{auzinsh_optically_2010,sobelman_introduction_1972}. Although the information in this subsection is not new, it is useful for the next subsection, where we will estimate analogous contributions due to nuclear tensor polarizability.
	
Let $E_{\eta I}$ be a nuclear energy corresponding to a state $|\eta IM\rangle$. Here, $I$ is the nuclear spin, $M$ is the magnetic quantum number and $\eta$ denotes all other relevant quantum numbers. By definition, the quadrupole moment of the nucleus is a second-rank tensor of the form
\begin{equation}
	Q_{ij}=\frac{3Q}{2I(2I-1)} \left(I_i I_j+I_j I_i-\frac{2}{3}I(I+1)\delta_{ij}\right)\,,\label{eq:Qik}
\end{equation}
where $Q$ is the expectation value of the electric quadrupole operator 
$\hat Q_{ij} = e(3r_i r_j - r^2 \delta_{ij})$ in the stretched state $|\eta I,M=I\rangle$. Here, $e$ is the charge of the proton. This quantity $Q$ may be related to the intrinsic nuclear electric quadrupole moment $Q_0$ (the quadrupole moment in the rotating body frame) via \cite{sobelman_introduction_1972,bohrmottleson}
\begin{equation}
	Q=\frac{I(2I-1)}{(I+1)(2I+3)}Q_0\,.\label{eq:Q}
\end{equation}
The intrinsic quadrupole moment $Q_0$ may, in turn, be related to the nuclear radius $R_0$ and the nuclear quadrupole deformation parameter $\beta_2$ \cite{ring_nuclear_2004} via
\begin{equation}
	Q_0=\frac{3}{\sqrt{5\pi}} eZ R_0^2 \beta_2\,.
	\label{Qbeta2}
\end{equation}
The values of quadrupole moments for different nuclei are tabulated in Ref.~\cite{stone_table_2005}. 
In the nuclear droplet model, the nuclear radius is described by the formula
\begin{equation}\label{R0}
	R_0=1.2\,A^{1/3}\,{\rm fm}\,.
\end{equation}
	
Within a model where the nucleus behaves as a deformed three-dimensional harmonic oscillator with frequencies $\omega_x=\omega_y\ne \omega_z$, one may derive an alternative representation for the quadrupole moment (\ref{Qbeta2}) \cite{Flambaum2016}:
\begin{equation}\label{Q_0}
	Q_0=\frac{2}{5}
	e Z R_0^2\bar{\omega}^2 \left(\frac{1}{\omega_z^2}-\frac{1}{\omega_x^2}\right),
\end{equation}
where $\bar{\omega}= \frac13( \omega_x + \omega_y+ \omega_z$) is a mean frequency which may be estimated using the phenomenological formula \cite{bohrmottleson}
\begin{equation}\label{baromega}
	\bar\omega = 41 A^{-1/3}\,{\rm MeV}\,.
\end{equation}
	
In an atom, the nuclear quadrupole moment is known to contribute to the hyperfine energy level splitting \cite{sobelman_introduction_1972, auzinsh_optically_2010} as
\begin{equation}
	\Delta\varepsilon_{\mathrm{hfs},Q}=B_Q\frac{\frac{3}{2}K(K+1)-2I(I+1)J(J+1)}{2I(2I-1)2J(2J-1)}\,,\label{eq:Ehfs}
\end{equation}
where $J$ is the electronic total angular momentum quantum number, $F$ is the atomic total angular momentum quantum number, which is the vector sum of $I$ and $J$, and $K\equiv F(F+1)-I(I+1)-J(J+1)$. The coefficient $B_Q$ is proportional to the nuclear quadrupole moment $Q$ and the expectation value of $1/r^3$ calculated with electronic radial wave functions,
\begin{equation}\label{BQ7}
	B_Q=eQ\left<r^{-3}\right> C_{IJ}\ .
\end{equation}
Here $C_{IJ}$ is a coefficient which takes into account the integration over angular variables. We do not specify the explicit value of this coefficient here, as it will drop out from the final result.
	
\subsection{Scalar and tensor nuclear polarizabilities}
	
When an external electric field $\boldsymbol{\mathcal{E}}$ is applied to a nucleus, the nuclear energy levels $E_{\eta I}$ are shifted due the quadratic Stark effect. These shifts may be written as $\Delta E_{\eta I} = -\frac12\alpha_{ij}{\cal E}_i{\cal E}_j$ where $\alpha_{ij}$ is, by definition, the electric nuclear polarizability tensor,
\begin{equation}
	\alpha_{ij} \equiv -2\sum_{\eta' I' M'}
	\frac{\langle \eta IM |  d_i | \eta' I'  M' \rangle \langle \eta' I' M'| d_j | \eta IM\rangle}{E_{\eta I} - E_{\eta' I'}}\,.
	\label{alpha-def}
\end{equation}
Here ${\bf d} = \sum_{k=1}^A q_k {\bf r}_k$ is the nuclear electric dipole operator, $q_k$ is the nucleon charge which appears due to the recoil effect, $q_k= e N/A$ for proton and $q_k = -eZ/A$ for neutron.
	
The symmetric tensor $\alpha_{ij}$ may be decomposed into a trace, $\alpha_0 = \sum_{k} \alpha_{kk}/3$, and a traceless part, $\alpha^{({\rm T})}_{ij} = \alpha_{ij}-\delta_{ij}\sum_k \alpha_{kk}/3$. Conventionally, the components $\alpha_0$ and $\alpha^{({\rm T})}_{ij}$ are referred to as the {\it scalar} and {\it tensor polarizabilities}, respectively, see, e.g., Ref.~\cite{auzinsh_optically_2010}.
	
The scalar polarizability will be considered in Sec.~\ref{sec:heavy}. In this section, we will focus on the tensor part. Similarly to the nuclear quadrupole moment (\ref{eq:Qik}), $\alpha^{({\rm T})}_{ij}$ may be expressed in terms of the nuclear spin operator $\hat I$ as
\begin{equation}
	\alpha^{({\rm T})}_{ij} = \frac{3\alpha_2}{2I(2I-1)}
	\left(I_i I_j + I_j I_i - \frac23 I(I+1) \delta_{ij} 
	\right),
	\label{alpha-tensor}
\end{equation}
where the coefficient $\alpha_2$ is the value of $\alpha^{({\rm T})}_{zz}$ calculated with the stretched state $|\eta II\rangle$.
Note that Eq.~(\ref{alpha-tensor}) defines the tensor polarizability $\alpha_2$ in the laboratory frame. The corresponding value in the rotating body frame $\alpha_2^{(0)}$ is related to $\alpha_2$ via an equation similar to \eqref{eq:Q}
\begin{equation}
	\alpha_2=\frac{I(2I-1)\alpha_2^{(0)}}{(I+1)(2I+3)}\,.
\end{equation}
	
The quantity $\alpha_2^{(0)}$ may be estimated in the deformed three-dimensional harmonic oscillator model with frequencies $\omega_x=\omega_y\neq\omega_z$ which was used to derive Eq.~(\ref{Q_0}). In cartesian coordinates, the nuclear states have the form $|n_x n_y n_z\rangle\equiv|n_x\rangle |n_y\rangle |n_z\rangle$ where $n_{x,y,z}$ are the quantum numbers in respective directions. With these functions, the diagonal components of the electric nuclear polarizability (\ref{alpha-def}) read
\begin{equation}
	\alpha_{ii}^{(0)} = 2\sum_{k=1}^A q_k^2
	\sum_{n'=n\pm1} \frac{\langle n | r_{k,i} |n'\rangle 
		\langle n' | r_{k,i} |n \rangle}{E^{(i)}_{n'}-E^{(i)}_n}\,,
	\label{eq-alphaii}
\end{equation}
where $E_n^{(i)}=\omega_i(n+\frac12)$ are the harmonic oscillator energies. Note that for any particular $i$, $\omega_i$ is assumed to be the same for all nucleons. The computation of the matrix elements in Eq.~(\ref{eq-alphaii}) is elementary \cite{griffiths2018introduction}. Using the identity $\sum_{k=1}^A q_k^2 =e^2 NZ/A$, one finds
\begin{equation}\label{alpha_body}
\begin{aligned}
    \alpha_2^{(0)} &= \frac23\alpha^{(0)}_{zz}-\frac13(\alpha^{(0)}_{xx}+\alpha^{(0)}_{yy}) \\
	&= \frac{2NZ}{3A}\frac{e^2}{m_p} \left(
	\frac1{\omega_z^2} - \frac1{\omega_x^2}
	\right),
\end{aligned}
\end{equation}
where $m_p$ is nucleon mass.
	
Since the tensors (\ref{eq:Qik}) and (\ref{alpha-tensor}) have the same structure, at the atomic level, the operator (\ref{alpha-tensor}) produces hyperfine energy level splitting analogous to (\ref{eq:Ehfs}), but with the constant $B_Q$ replaced by
\begin{equation}\label{Ba14}
	B_\alpha=e^2\alpha_2\left<r^{-4}\right> C_{IJ}\,.
\end{equation}
An estimate of the shifts due to tensor nuclear polarizability may thus be obtained by computing the ratio $B_{\alpha}/B_Q$. Note that the power of $r$ in Eq.~\eqref{Ba14} is different from that in Eq.~\eqref{BQ7} because the operators \eqref{eq:Qik} and \eqref{alpha-tensor} have different dimensions. Making use of Eqs.~\eqref{Q_0} and \eqref{alpha_body}, we find
\begin{equation}
	\frac{B_\alpha}{B_Q}=\frac{5Ne^2}{3Am_p\bar{\omega}^2R_0^2}\frac{\left< r^{-4} \right>}{\left< r^{-3}\right>}\,.
	\label{BBQ}
\end{equation}
	
The expectation values of the operators $r^{-3}$ and $r^{-4}$ in Eq.~(\ref{BBQ}) receive main contributions from the near-nucleus region, where $r\ll a_{ B}/Z^{1/3}$. In this region, the screening of the Coulomb field of the nucleus is negligible and the electron radial wave functions may be well approximated by the Bessel functions \cite{khriplovich_parity_1991},
\begin{eqnarray}
	f_{njl}(r)&=& \frac{c_{njl}}r
	\left[ (\gamma+\kappa)J_{2\gamma}(x) - \frac x2 J_{2\gamma-1}(x) \right],
	\nonumber\\
	g_{njl}(r) &=& \frac{c_{njl}}{r}Z\alpha J_{2\gamma}(x)\,,
	\label{Bessel-functions}
\end{eqnarray}
where $x\equiv\sqrt{8Zr/a_{B}}$, $\gamma=\sqrt{\kappa^2-Z^2\alpha^2}$ and $\kappa=(-1)^{j-l+1/2}(j+1/2)$. The value of the normalization constant $c_{njl}$ may be found in Ref.~\cite{khriplovich_parity_1991}. For our purpose, the explicit value of this constant is not needed since it cancels out in the ratio \eqref{BBQ}. 
	
With the wave functions (\ref{Bessel-functions}), the expectation value of the $r^{-p}$ operator may be found analytically for $p<1+2\gamma$,
\begin{equation}
	\begin{aligned}
		\frac{\langle r^{-p} \rangle}{c_{njl}^2} &=\frac1{32}\left( \frac{8Z}{a_{\rm B}} \right)^{p-1}
		\frac{\Gamma(p-3/2)\Gamma(1+2\gamma-p)}{\sqrt\pi \Gamma(p)\Gamma(p+2\gamma)}\\ 
		&\times \Big[
		2\gamma^2 +p(5+p(p-4)+4Z^2\alpha^2) - 6Z^2\alpha^2\\
		&
		+ \kappa^2(4p-6)
		-2\kappa(p-1)(2p-3) -2
		\Big].
		\label{r-p-exp}
	\end{aligned}
\end{equation}
Here we have used the wave functions (\ref{Bessel-functions}) with $j>1/2$ because the expectation values of the operators (\ref{eq:Qik}) and (\ref{alpha-tensor}) vanish in $s_{1/2}$ and $p_{1/2}$ states. The equation (\ref{r-p-exp}) allows us to estimate the ratio of the expectation values  operators $r^{-4}$ and $r^{-3}$ for heavy atoms ($Z>70$) in the $p_{3/2}$ state
\begin{equation}
	\frac{\langle r^{-4} \rangle}{\langle r^{-3} \rangle}
	= \frac{\xi Z}{a_B} \,, \quad \xi\approx 1.3\,.
	\end{equation}
Substituting this relation into Eq.~(\ref{BBQ}), we find
\begin{equation}\label{ratio-result}
	\left|\frac{B_\alpha}{B_Q}\right| = \frac{5\xi N Ze^2}{3A a_B m_p\bar{\omega}^2R_0^2}\approx6.1\times10^{-7}Z\,,
\end{equation}
where we have used Eqs.~\eqref{R0} and \eqref{baromega} and assumed the approximations $N\approx 1.5 Z$ and $A\approx2.5Z$ for heavy nuclei.
	
Numerically, for heavy atoms with $Z\sim 100$, the ratio in Eq.~(\ref{ratio-result}) is on the order of $10^{-4}$ and is smaller for lighter elements. The effect of the tensor nuclear polarizability on electronic spectra is nearly four orders in magnitude smaller than that of the quadrupole nuclear moment. As a result, in many cases, the nuclear tensor polarizability effect may be neglected. In the next section, we will focus on the effects of scalar nuclear polarizability.

\section{Energy shift due to scalar nuclear polarizability in medium and heavy hydrogen-like ions}\label{sec:heavy}
	
In this section, we study the energy level shifts in medium and heavy hydrogen-like ions due the nuclear polarization induced by the electron-nucleon interaction. In Sec.~\ref{theor}, we review the necessary theoretical background developed in the papers \cite{plunien_nuclear_1989,plunien_nuclear_1991,plunien_nuclear-polarization_1995,plunien_erratum:_1996,nefiodov_nuclear_1996}. In the subsequent subsections, we present and discuss the results of numerical calculations of these effects.
	
\subsection{Theoretical background}\label{theor}
	
It is well known that nuclear polarization due to the electron-nucleon interaction contributes to the electronic Lamb shift, see, e.g., Ref.~\cite{Mohr-review} for a review. In light atoms and ions this effect is very small \cite{pachucki_nuclear-structure_1993,pachucki_theory_1994}, but it becomes significant in heavy atoms and ions, where the electron wave functions have sizable overlap with the nucleus and may thus be considerably affected by the nuclear structure. In this section we study these energy shifts in hydrogen-like medium and heavy ions. The effects in multieletron atoms will be discussed in Sec.~\ref{Eff_Pot_Sec}. Motivated by future study of the nonlinearity of King's plot \cite{king_isotope_2013} induced by the nuclear polarization, we will mainly focus on the even-even nuclei with vanishing nuclear spin.
	
\begin{figure}[tbh]
	\centering
	\includegraphics[width=5cm]{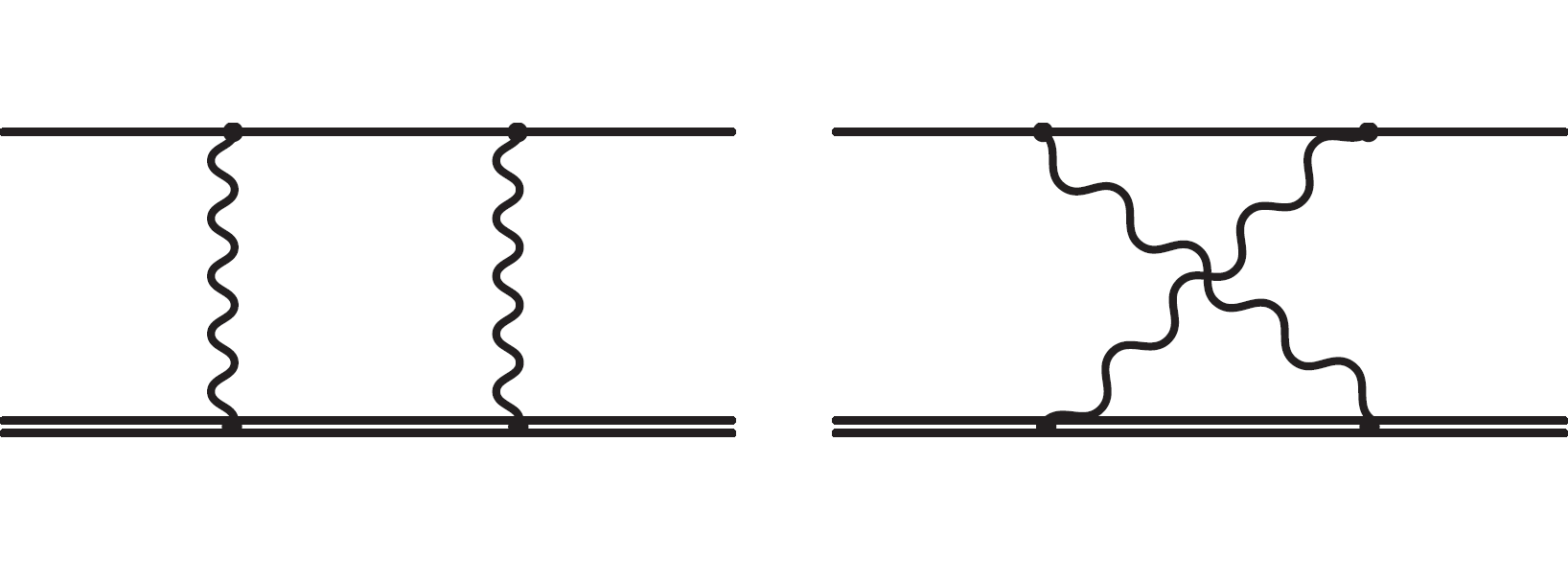}
	\caption{Contributions to electronic self energy due to nuclear polarization. Solid and double solid lines correspond to the electron and nucleon propagators, respectively, while the wavy lines represent the photon propagators.}
	\label{fig1}
\end{figure}

The atomic energy shifts due to nuclear polarization are well-understood within the framework of QED and may be represented by a two-photon exchange between the electron and an unpaired nucleon in the nucleus. This process may be illustrated by the Feynman diagrams in Fig.~\ref{fig1}. 
	
As mentioned above, we are considering nuclei with vanishing angular momentum in the ground state, $L=0$. The excited nuclear states may, on the other hand, have arbitrary angular momentum $L$ and energy $E_L$. The initial electronic state may be characterised by its principal quantum number $n$, its orbital angular momentum $l$ and its total angular momentum $j$. The electronic energies will be denoted by $\varepsilon_{nlj}$. The atomic energy level shift due to the processes presented in Fig.~\ref{fig1} was calculated in Ref.~\cite{nefiodov_nuclear_1996} and reads
\begin{widetext}
	\begin{equation}
		\begin{aligned}
			\Delta \varepsilon_{nlj}&=-\frac{\alpha}{4\pi}(2L+1)B(EL;L\rightarrow 0)\sum_{j'}(2j'+1)\begin{pmatrix} j' & j & L \\ \frac{1}{2} & -\frac{1}{2} & 0\ \end{pmatrix}^2\\ 
			&\times \left(\sum_{n'l'} \frac{|\langle nlj|F_L|n'l' j' \rangle|^2}{\varepsilon_{n'l'j'}-\varepsilon_{nlj}+E_L} 
			+
			\int_{-\infty}^{-{m_e}} \frac{|\langle nlj|F_L|\varepsilon j' \rangle|^2}{\varepsilon-\varepsilon_{nlj}-E_L}d\varepsilon
			+\int_{m_e}^\infty \frac{|\langle nlj|F_L|\varepsilon j' \rangle|^2}{\varepsilon-\varepsilon_{nlj}+E_L}d\varepsilon 
			\right)\,,
			\label{main-formula}			\end{aligned}
	\end{equation}
\end{widetext}
where $B(EL;L\rightarrow 0)$ is the reduced transition probability for nuclear electric transitions from an excited state with angular momentum $L$ to the ground state and $F_L(r)$ is a radial function of the form
\begin{equation}\label{radial-function}
	\begin{aligned}
		F_L(r)&=\frac{4\pi}{(2L+1)R^L_0}\left(\frac{r^L}{R_0^{L+1}}\Theta(R_0-r)  \right.\\
		&\left. + \frac{R_0^L}{r^{L+1}}\Theta(r-R_0) \right) \,,\quad (L\geq 1)\,.
	\end{aligned}
\end{equation}
which behaves like $1/r^{L+1}$ outside the nucleus and like $r^L$ inside. The function (\ref{radial-function}) represents a regularization of the $1/r^{L+1}$ potential to the case of an extended nucleus. 
	
The three terms in the second line in Eq.~(\ref{main-formula}) correspond to the contributions from intermediate electronic states in discrete, negative energy continuum and positive energy continuum spectra, respectively. The matrix elements in these terms are defined as
\begin{equation}\label{radial-integral}
	\langle A|F_L|B \rangle \equiv \int_0^\infty dr\, r^2 F_L(f_Af_B + g_Ag_B)\,,
\end{equation}
where $f_{A,B}$ and $g_{A,B}$ are, respectively, upper and lower Dirac radial wave functions of the electron.
	
It was pointed out in Refs.~\cite{plunien_nuclear_1991,plunien_nuclear-polarization_1995} that the intermediate electronic states in the discrete spectrum give negligible contribution to the energy shift as compared to the contributions from the lower and upper continua. Indeed, the radial integral in Eq.~(\ref{radial-integral}) receives the main contribution from the vicinity to the surface of the nucleus, where the radial function (\ref{radial-function}) peaks. Thus, the dominant contributions come from small distances, i.e., from states with high energies in the continuous spectrum. Moreover, the discrete spectrum terms in Eq.~(\ref{main-formula}) are suppressed by large denominators because 
$\varepsilon_{n'l'j'}-\varepsilon_{nlj}\ll E_L$. Therefore, in our calculations below, we will ignore the contributions from the discrete spectrum.
	
In Eq.~(\ref{main-formula}), the leading contributions come from the low-$L$ transitions while the higher-$L$ terms are suppressed because the corresponding electron wave functions have a small overlap with the nucleus. Therefore, to a good degree of accuracy, it is sufficient to consider only the terms with $L=1$ and $L=2$, the former corresponds to a nuclear giant electric dipole resonance while the latter may be interpreted as a contribution from nuclear rotation associated with the collective nuclear quadrupole moment in a deformed nucleus. In the following subsections, we will consider these two contributions separately. 
	
\subsection{Contribution from giant electric dipole resonance transition}\label{Ions_E1}
	
The giant electric dipole resonance nuclear transitions correspond to $L=1$ in Eq.~(\ref{main-formula}),
	\begin{equation}
\begin{aligned}
		&\Delta \varepsilon_{nlj}=-\frac{3\alpha}{4\pi}B(E1)
		\sum_{j'}(2j'+1)\begin{pmatrix} j' & j & 1 \\ \frac{1}{2} & -\frac{1}{2} & 0\ \end{pmatrix}^2
		\\&\times\left(
		\int_{-\infty}^{-{m_e}} \frac{|\langle nlj|F_1|\varepsilon j' \rangle|^2 d\varepsilon}{\varepsilon-\varepsilon_{nlj}-E_{\rm GR}}
		+ \int_{m_e}^\infty \frac{|\langle nlj|F_1|\varepsilon j' \rangle|^2d\varepsilon}{\varepsilon-\varepsilon_{nlj}+E_{\rm GR}}\right),\label{main-formula-E1}
\end{aligned}
	\end{equation}
where the energy of giant dipole resonance $E_{\rm GR}$ in a heavy nucleus is given by \cite{hoffmann1984effects,ring_nuclear_2004}:
\begin{equation}\label{EGR}
	E_\mathrm{GR} = 95(1-A^{-1/3})A^{-1/3}\ \mathrm{MeV}\,.
\end{equation}
	
The transition probability $B(E1)\equiv B(E1;1\rightarrow 0)$ in Eq.~(\ref{main-formula-E1}) for giant electric dipole resonance transitions may be estimated using the Thomas-Reiche-Kuhn sum rule \cite{ring_nuclear_2004}, giving
\begin{equation}\label{nuc-strength}
	B(E1)\equiv B(E1;1\rightarrow0)=\frac{3}{8\pi}\frac{Z(A-Z)e^2}{AE_\mathrm{GR}m_p}\,.
\end{equation}
Using the formula (\ref{main-formula-E1}) with the nuclear transition strength (\ref{nuc-strength}) we numerically calculate the energy shifts for $1s$, $2s$ and $2p_{1/2}$ states in hydrogen-like ions. The radial integrals (\ref{radial-integral}) are calculated numerically using known continuum and discrete state Dirac radial wave functions, taking into account the finite size of the nucleus. The integrals over $d\varepsilon$ in Eq.~(\ref{main-formula-E1}) is also evaluated numerically. 
	
We consider hydrogen-like ions with nuclear charges ranging from $Z=20$ (Calcium) to $Z=98$ (Californium), and extend the results to superheavy elements up to $Z=136$. For each $Z$, ions with different $A$ are also considered. It should be noted that although we restrict ourselves to even-even nuclei, Eq.~\eqref{main-formula-E1} also applies to nuclei with odd $A$, only in this case the structure of atomic energy levels is more complicated due to the hyperfine interactions.  
On the other hand, extending the current computation to odd-$A$ nuclei proves to be convenient for fitting the results, see Eqs.~\eqref{A0} and \eqref{Delta_A_Def} below.
	
The results of our numerical calculation are presented in Table~\ref{Results_Ions_E1}. In Table \ref{tab:plunien_vs_we}, we compare our results with those published earlier \cite{plunien_nuclear-polarization_1995,plunien_nuclear_1991} for certain heavy elements. This table shows that our numerical methods provide the accuracy within 5\% of earlier publications~\cite{plunien_nuclear-polarization_1995,plunien_nuclear_1991}.

%Table II
\begin{center}
	\begin{table}[htb]
		\begin{tabular}{cc| cc  | cc  | cc}
			\hline\hline
			$Z$ & $A$ &  $\Delta \varepsilon_{1s}$ & $\Delta \varepsilon_{1s}^\mathrm{ref}$ & $\Delta \varepsilon_{2s}$ & $\Delta \varepsilon_{2s}^\mathrm{ref}$ & $\Delta \varepsilon_{2p}$ & $\Delta \varepsilon_{2p_{1/2}}^\mathrm{ref}$ \\
			\hline
			82	&	208	&	$	18.2	$	&	$	17.3	$	&	$	3.2	$	&	$	3.0	$	&	$	0.27$	&	$	-	$	\\
			90	&	232	&	$	39.6	$	&	$	41.8	$	&	$	7.4	$	&	$	7.7	$	&	$	0.80$	&	$	0.84	$	\\
			92	&	234	&	$	47.5	$	&	$	49.8	$	&	$	9.1	$	&	$	9.4	$	&	$	1.0	$	&	$	1.1	$	\\
			92	&	236	&	$	47.8	$	&	$	50.1	$	&	$	9.1	$	&	$	9.4	$	&	$	1.1	$	&	$	1.1	$	\\
			92	&	238	&	$	48.1	$	&	$	42.4	$	&	$	9.2	$	&	$	8.1    $&	$	1.1	$	&	$	1.0	$	\\
			98	&	250	&	$	84.8	$	&	$	87.2	$	&	$	17.2$	&	$	17.3	$&	$	2.4	$	&	$	2.4	$	\\
			98	&	252	&	$	85.3	$	&	$	87.6	$	&	$	17.3$	&	$	17.4	$&	$	2.4	$	&	$	2.5	$	\\
			\hline\hline
		\end{tabular}
		\caption{Comparison of our calculations for the nuclear giant dipole resonance contributions to the energy shifts due to nuclear polarization \eqref{main-formula-E1} with earlier calculated values presented in Refs.~\cite{plunien_nuclear-polarization_1995} and \cite{nefiodov_nuclear_1996}. The values of $\Delta\varepsilon$ are given in units of meV.}\label{tab:plunien_vs_we}
	\end{table}  
\end{center}

\begin{center}
		\begin{figure*}[htb]
			\centering
			\includegraphics[scale=0.28]{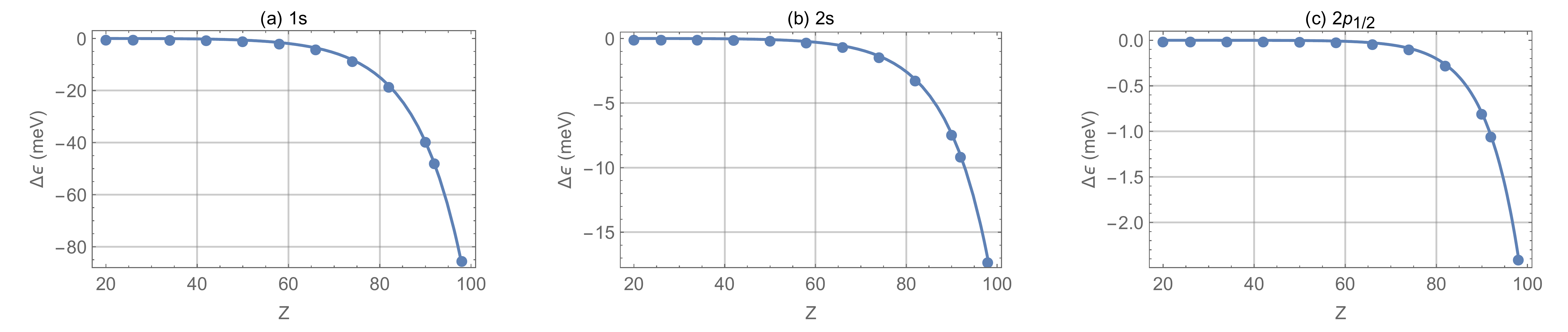}
			\caption{Shifts of $1s$, $2s$ and $2p_{1/2}$ energy levels in medium and heavy hydrogen-like ions (dots) due to nuclear polarization through E1 nuclear transitions.
			Solid lines represent the best fit of these shifts with the function (\ref{eq:fit}).}
			\label{Plot_Results_Ions_E1}
		\end{figure*}
	\end{center}

One of goals in this paper is to establish an analytical dependence $\Delta \varepsilon=\Delta \varepsilon(Z,A)$, which should give the energy shifts for medium and  heavy ions including isotopic dependence. We find that to a high degree of accuracy, the isotopic dependence of the energy shifts is linear. For each $Z$ we consider the isotope with the atomic number\footnote{In Eq.~(\ref{A0}), the square brackets denote the rounding to the nearest integer. This formula covers most of the stable isotopes (when exist) for each $Z$ in the region $20\leq Z\leq 98$. We note also that the choice of $A_0$ here is entirely for convenience. Different choices require different fitting functions and parameters but are otherwise equivalently legitimate.} 
\begin{equation} \label{A0}
	A_0(Z)=[-17.62+2.737 Z]\,,
\end{equation} 
and write the mass number of the other isotopes as 
\begin{equation}\label{Delta_A_Def}
	A=A_0(Z)+\Delta A\,.
\end{equation}
In the series decomposition of energy shift $\Delta \varepsilon$, it is sufficient to keep only linear terms in $\Delta A$, $\Delta\varepsilon(Z,A)=\Delta\varepsilon_0(Z)+\Delta\varepsilon_1(Z)\Delta A$. The functional behaviour of $\Delta\varepsilon_0(Z)$ is presented by dots in Fig.~\ref{Plot_Results_Ions_E1} for $1s$, $2s$ and $2p_{1/2}$ states. These graphs show that $-\Delta\varepsilon_0(Z)$ grows approximately exponentially with $Z$. It may be verified that $\Delta\varepsilon_1(Z)$ shows similar behaviour. Therefore, we use the following fitting functions to approximate the energy shift,
\begin{equation}
	\begin{aligned}\label{eq:fit}
		\Delta \varepsilon(Z,\Delta A) &=-\big[\exp(a_0+a_1Z+a_2Z^2)Z^{a_3}\\
		&+\exp(b_0+b_1Z+b_2Z^2)Z^{b_3}\Delta A\big]\,{\rm meV}\,,
	\end{aligned} 
\end{equation}
with fitting parameters $a_{0,1,2,3}$ and $b_{0,1,2,3}$ presented in  Table~\ref{Table_for_c}.  With these parameters, equation (\ref{eq:fit}) reproduces the numerical results in Table~\ref{Results_Ions_E1} with an accuracy under 2\% for $20\leq Z\leq 98$. As a demonstration, the functions (\ref{eq:fit}) are plotted in Fig.~\ref{Plot_Results_Ions_E1} for $1s$, $2s$ and $2p_{1/2}$ states.

In computing the energy level shifts in Table~\ref{Results_Ions_E1}, we employed the empirical formula (\ref{R0}) for the nuclear radius. This formula is, however, only approximate, and experimental values of $R$ may have deviations from Eq.~(\ref{R0}),
\begin{equation}\label{radius-deviation}
	R=R_0 + \delta R\,.
\end{equation}
To take into account such deviations, we modify Eq.~\eqref{eq:fit} as
\begin{equation}
	\Delta \varepsilon(Z,A,R) = \Delta \varepsilon(Z,A,R_0) + \delta_R \varepsilon(Z) \delta R\,,
	\label{correction-radius}
\end{equation}
where $\Delta \varepsilon(Z,A,R_0)$ is given by Eq.~(\ref{eq:fit}), and the correction term $\delta_R \varepsilon(Z)$ may be approximated by the function
\begin{equation}\label{c0c1c2}
	\delta_R \varepsilon(Z) = \exp(c_0 + c_1 Z + c_2 Z^2 )Z^{c_3}\,{\rm meV}/{\rm fm}\,.
\end{equation}
%Table III
\begin{center}
	\begin{table}[tbh]
		\begin{tabular}{c|c|c|c}
			\hline\hline
			& $1s$                           & $2s$                  & $2p_{1/2}$             \\
			\hline
			$a_0$    & $-22.2$               & $-24.4$               & $-36.3$                \\
			$a_1$    & $-2.21\times 10^{-2}$ & $-2.36\times 10^{-2}$ & $-2.92\times 10^{-2}$  \\
			$a_2$    & $3.07\times 10^{-4}$  & $3.66\times 10^{-4}$  & $4.40\times 10^{-4}$   \\
			$a_3$    & 5.64                  & 5.67                  & 7.80                   \\
			\hline
			$b_0$    & $-19.9$               & $-22.0$               & $-33.5$                \\
			$b_1$    & $4.23\times 10^{-3}$  & $3.44\times 10^{-3}$  & $4.41\times 10^{-3}$   \\
			$b_2$    & $1.77\times 10^{-4}$  & $2.32\times 10^{-4}$  & $2.82\times 10^{-4}$   \\
			$b_3$    & 3.56                  & 3.57                  & 5.52                   \\
			\hline
			$c_0$    & $-24.0$               & $-26.2$               & $-38.1$                \\
			$c_1$    & $-1.70\times 10^{-2}$ & $-1.94\times 10^{-2}$ & $-2.56\times 10^{-2}$  \\
			$c_2$    & $3.39\times 10^{-4}$  & $4.02\times 10^{-4}$  & $4.75\times 10^{-7}$   \\
			$c_3$    & 5.44                  & 5.49                  & 7.63                   \\
			\hline\hline
		\end{tabular}
		\caption{Numerical values of the fitting parameters in the formulae \eqref{eq:fit} and \eqref{c0c1c2} describing $1s$, $2s$ and $2p_{1/2}$ energy level shifts in hydrogen-like ions due to nuclear giant electric dipole resonance.}\label{Table_for_c}
	\end{table}
\end{center}
	
The coefficients $c_{0,1,2,3}$ in Eq.~(\ref{c0c1c2}) are computed numerically and collected in Table~\ref{Table_for_c}. We stress that the corrections due to variations of the nuclear radius are important within the study of possible nonlinearity of King's plot \cite{king1963,king_isotope_2013,Gebert2015,Knollmann2019,Manovitz2019} and physics beyond the SM \cite{Frugiuele2017,berengut2018,flambaum_isotope_2018,Yerokhin2020,berengut2020,berengut_generalized_2020}. 
	
We note that here we consider only the giant dipole resonance nuclear transitions with the nuclear energy (\ref{EGR}) and reduced transition probability (\ref{nuc-strength}). In certain nuclei, such as ${}^{228}$Th, there can be additional E1 transitions from the ground state to low lying levels with typical energy in the keV range. As a result, the contributions of such transitions to the overall energy shift are, to a certain degree, enhanced when compared to other typical transitions with energy in the MeV range. However, we point out that the probability $B(E1)$ of these low lying transitions is three to four orders of magnitude smaller than that of the giant dipole resonance. Therefore, contributions from isolated low-lying E1 nuclear energy levels are still negligible.

Finally, we point out that the accuracy of calculation of the energy shift with Eq.~(\ref{main-formula-E1}) strongly depends on the value of the nuclear reduced transition probability (\ref{nuc-strength}). The latter formula provides an approximate, average description of giant dipole resonance transitions, and particular isotopes may have considerable deviations from this formula. For such isotopes one can improve the accuracy of calculations of the energy shift by applying the correcting coefficient $B_{\rm exact}(E1)/B(E1)$, where $B_{\rm exact}(E1)$ is the exact value of the nuclear reduced transition probability found from experiments and $B(E1)$ is the approximate value calculated with the use of Eq.~(\ref{nuc-strength}). In particular, the values of $B_{\rm exact}(E1)$ may be derived from the photonuclear cross-section data collected, e.g., in Ref.~\cite{Kawano:2019tsn}.

\subsection{Contribution from nuclear rotational transition}\label{Ions_E2}
	
A spinless nucleus with quadrupole deformation may have collective E2 excitations from the ground state into the rotational band. The energies of these transitions are typically on the order of a few dozens keV, that is, much lower than the energy of giant electric dipole resonance transition which is about a dozen of MeV. As a result, the shifts due to nuclear rotational transitions receive sizable enhancements.
	
It is worth noting that nuclear transitions to higher rotational states give minor contributions to the atomic energy level shifts \cite{plunien_nuclear-polarization_1995},
because the reduced transition probability $B(E2;2\rightarrow 0)$ decreases rapidly for such states, and there is additional suppression from higher nuclear energy in the denominator.
Therefore, in this paper we consider only the nuclear transition from the ground state to the lowest rotational state with $L=2$. In this case, Eq.~(\ref{main-formula}) may be written as
	\begin{equation}
\begin{aligned}
	&\Delta \varepsilon_{nlj}=-\frac{5\alpha}{4\pi}B(E2)
		\sum_{j'}(2j'+1)\begin{pmatrix} j' & j & 2 \\ \frac{1}{2} & -\frac{1}{2} & 0\ \end{pmatrix}^2
	\\&
		\times\left(\int_{-\infty}^{-{m_e}} \frac{|\langle nlj|F_2|\varepsilon j' \rangle|^2d\varepsilon}{\varepsilon-\varepsilon_{nlj}-E_{\rm rot}} +\int_{m_e}^\infty \frac{|\langle nlj|F_2|\varepsilon j' \rangle|^2d\varepsilon}{\varepsilon-\varepsilon_{nlj}+E_{\rm rot}}\right),
\end{aligned}
		\label{main-formula-E2}
	\end{equation}
where $E_{\rm rot}$ is the energy of the lowest nuclear rotational level. 
	
The transition probability $B(E2)\equiv B(E2;2\rightarrow0)$ may be expressed via the intrinsic nuclear quadrupole moment \eqref{Q_0} as  $B(E2;2\rightarrow0)=Q_0^2/(16\pi)$ (see, e.g., Ref.~\cite{ring_nuclear_2004}), which gives
\begin{equation}
	B(E2)\equiv B(E2;2\rightarrow 0)=\frac{1}{5}\left(\frac{3}{4\pi}\right)^2 Z^2e^2R_0^4 \beta_2^2\,.
	\label{BE2}
\end{equation}
Numerical values of the nuclear deformation parameter $\beta_2$ as well as the reduced transition probability $B(E2)$ may be found in, e.g., Ref.~\cite{Raman:1201zz}. In this reference, the values of the reduced transition probability are calculated from known values of the lifetime of the excited nuclear state. Lifetimes of the excited nuclear states may be found, e.g., in \cite{NuDat}.

According to an empirical rule \cite{ring_nuclear_2004}, the energy of the first excited $2^+$ nuclear rotational state is connected to its transition probability from the ground state via
\begin{equation}
	E_{\rm rot} B(E2)\approx 25Z^2 A^{-1} e^2\ \mathrm{MeV}\times\mathrm{fm}^4\,,
\end{equation}
which, along with Eq.~\eqref{BE2}, allow us to express the energy $E_{\rm rot}$ in terms of macroscopic nuclear parameters,
\begin{equation}\label{rot-energy}
	E_{\rm rot}\approx \frac{2\pi^2}{9 A R_0^4\beta_2^2} \mathrm{GeV}\times\mathrm{fm}^4\,.
\end{equation}
It is worth noting that this formula is applicable to deformed nuclei with $0.2\lesssim \beta_2\lesssim0.35$.
	
Substituting Eqs.~(\ref{BE2}) and (\ref{rot-energy}) into Eq.~(\ref{main-formula-E2}), one may calculate numerically the atomic energy level shifts due to nuclear rotational transitions using a procedure analogous to that for the giant electric dipole resonance case. There is, however, one important feature in the rotational case: The reduced transition probability (\ref{BE2}) depends explicitly on the nuclear deformation parameter $\beta_2$ which is, in general, a nonmonotonic function of $Z$ and $A$. Moreover, experimental data of $B(E2)$ collected, e.g., in Ref.~\cite{Raman:1201zz} have some deviation from formula \eqref{BE2}. As a result, we shall keep $B(E2)$ as a free parameter and calculate the quantity $\Delta\varepsilon_{nlj}/B(E2)$, i.e., the electronic matrix element only. 
	
Another source of dependence on $\beta_2$ is the rotational energy $E_{\rm rot}$ in the denominator in Eq.~\eqref{main-formula-E2}. However, the energy integrals in Eq.~\eqref{main-formula-E2} vary slowly for different values of $E_{\rm rot}$ given by Eq.~\eqref{rot-energy} because these integrals receive dominant contributions from $\varepsilon\approx50m_e$ whereas $E_{\rm rot}\lesssim 1$ MeV. Thus, in our numerical calculation, we use an average value $\beta_2=0.27$ to estimate the rotational energy $E_{\rm rot}$ in the denominator in Eq.~\eqref{main-formula-E2}.

We numerically calculate the quantity $\Delta\varepsilon_{nlj}/B(E2)$ for $1s$, $2s$ and $2p_{1/2}$ energy levels in  hydrogen-like ions with nuclear deformations $0.2\lesssim \beta_2\lesssim0.35$. The results of these calculations are collected in Table~\ref{Results_Ions_E2}. Comparisons with known results are presented in Table \ref{tab:plunien_vs_weE2}.

\begin{center}
\begin{figure*}[tbh]
			\centering
			\includegraphics[scale=0.28]{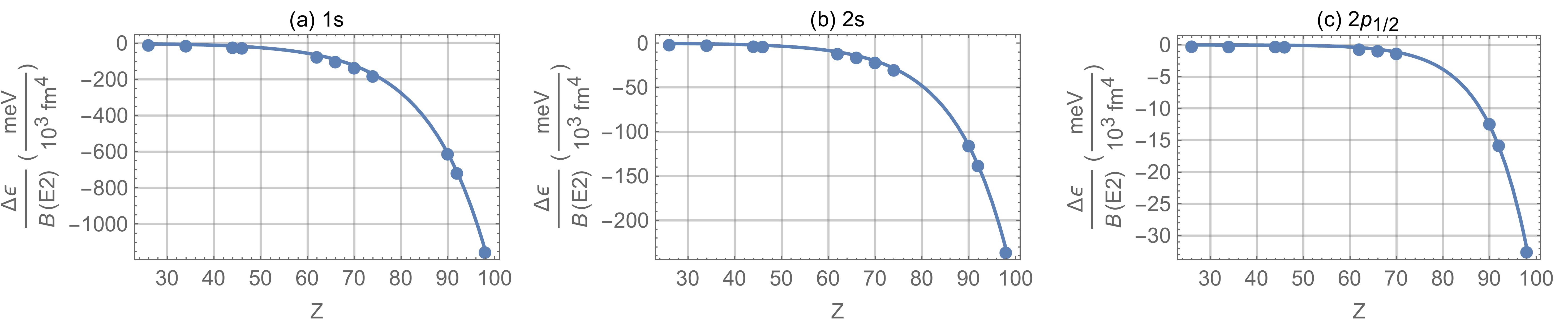}
			\caption{Shifts of $1s$, $2s$ and $2p_{1/2}$ energy levels in heavy hydrogen-like ions (dots) due to nuclear polarization through rotational E2 transitions.
			Solid lines represent the best fit of these shifts with the function (\ref{eq:fitE2}).}
\label{Plot_Results_Ions_E2}
\end{figure*}
\end{center}

%Table V
\begin{center}
	\begin{table}[tbh]
		\begin{tabular}{cc| cc  | cc  | cc}
			\hline\hline
			$Z$ & $A$ &  $\frac{\Delta \varepsilon_{1s}}{B(E2)}$ & $\frac{\Delta \varepsilon_{1s}^\mathrm{ref}}{B(E2)}$ & $\frac{\Delta \varepsilon_{2s}}{B(E2)}$ & $\frac{\Delta \varepsilon_{2s}^\mathrm{ref}}{B(E2)}$ & $\frac{\Delta \varepsilon_{2p}}{B(E2)}$ & $\frac{\Delta \varepsilon_{2p}^\mathrm{ref}}{B(E2)}$ \\
			\hline
			90	&	230	&	$	0.602	$	&	$	0.605	$	&	$	0.114	$	&	$	0.112	$	&	$	0.0122    $	&	$	0.0120	$	\\
			92	&	234	&	$	0.710   $	&	$	0.708	$	&	$	0.136	$	&	$	0.133	$	&	$	0.0156	$	&	$	0.0153	$	\\
			92	&	236	&	$	0.705	$	&	$	0.702	$	&	$	0.135	$	&	$	0.132	$	&	$	0.0155	$	&	$	0.0152	$	\\
			98	&	250	&	$	1.15	$	&	$	1.12	$	&	$	0.235   $	&	$	0.213	$&	$	0.0324	$	&	$	0.0311	$	\\
			98	&	252	&	$	1.14	$	&	$	1.07	$	&	$	0.234   $	&	$	0.213	$&	$	0.0322	$	&	$	0.0298	$	\\
			\hline\hline
		\end{tabular}
		\caption{Comparison of our calculations for the nuclear rotational transition contributions to the energy shifts due to nuclear polarizability \eqref{main-formula-E2} with previously calculated values presented in Refs.~\cite{plunien_nuclear-polarization_1995} and \cite{nefiodov_nuclear_1996}. The quantities $\Delta\varepsilon/B(E2)$ are given here in units of ${\rm meV}/{\rm fm}^4$.
		\label{tab:plunien_vs_weE2}}
	\end{table}  
\end{center}

For each ion, we consider several isotopes to determine the isotopic dependence of the energy shift with respect to the quantity $\Delta A$ as defined in Eq.~\eqref{Delta_A_Def}. We find that these results may be approximated with the exponential function,
\begin{equation}
	\begin{aligned}
		&\frac{\Delta \varepsilon(Z,\Delta A)}{B(E2)}=  \big[-\exp(\tilde{a}_0+\tilde{a}_1Z+\tilde{a}_2Z^2)Z^{\tilde{a}_3} \\
		&+\exp(\tilde{b}_0+\tilde{b}_1Z+\tilde{b}_2Z^2)Z^{\tilde{b}_3}\Delta A\big]\, \frac{\rm meV}{10^3{\rm fm}^4}\,,
		\label{eq:fitE2}
	\end{aligned}
\end{equation}
where the values of the coefficients $\tilde{a}_{0,1,2,3}$ and $\tilde{b}_{0,1,2,3}$ are collected in Table~\ref{tab:coefsE2}. The function (\ref{eq:fitE2}) with $\Delta A=0$ is plotted in Fig.~\ref{Plot_Results_Ions_E2} for $1s$, $2s$ and $2p_{1/2}$ energy level shifts.
%Table VI
\begin{center}
	\begin{table}[htb]
		\begin{tabular}{c|c|c|c}
			\hline\hline
			                 & $1s$                 & $2s$                  & $2p_{1/2}$            \\
			\hline
			$\tilde{a}_0$	 & $-5.78$	            & $-10.3$	            & $-24.8$	            \\
			$\tilde{a}_1$	 & $-8.74\times10^{-3}$ & $-1.10\times 10^{-2}$	& $-1.92\times 10^{-2}$	\\
			$\tilde{a}_2$	 & $2.74\times 10^{-4}$	& $3.36\times 10^{-4}$	& $4.18\times 10^{-4}$  \\
			$\tilde{a}_3$    & 3.41	                & 3.46	                & 5.70	                \\
			\hline
			$\tilde{b}_0$	 & $-8.27$	            & $-12.7$	            & $-26.9$               \\
			$\tilde{b}_1$	 & $-1.59\times10^{-2}$ & $-1.80\times 10^{-2}$	& $-2.33\times 10^{-2}$ \\
			$\tilde{b}_2$	 & $3.02\times 10^{-4}$	& $3.64\times 10^{-4}$	& $4.33\times 10^{-4}$	\\
			$\tilde{b}_3$	 & 2.81	                & 2.85	                & 4.96                  \\
			\hline
			$\tilde{c}_0$	 & $-7.06$	            & $-11.5$	            & $-25.8$               \\
			$\tilde{c}_1$	 & $-1.54\times10^{-2}$	& $-1.78\times 10^{-2}$	& $-2.51\times 10^{-2}$ \\
			$\tilde{c}_2$	 & $3.00\times 10^{-4}$	& $3.64\times 10^{-4}$	& $4.41\times 10^{-4}$  \\
			$\tilde{c}_3$    & 3.52	                & 3.57	                & 5.75                  \\
			\hline\hline
		\end{tabular}
		\caption{Numerical values of the fitting coefficients in Eqs.~\eqref{eq:fitE2} and \eqref{eq:dRE2} describing $1s$, $2s$ and $2p_{1/2}$ energy level shifts in hydrogen-like ions due to nuclear E2 rotational transitions in deformed nuclei.}\label{tab:coefsE2}
	\end{table}	
\end{center}
	
Analogously to Eqs.~(\ref{correction-radius}) and (\ref{c0c1c2}), we find the corrections due to the deviation of the actual nuclear radius (\ref{radius-deviation}) from the approximate formula (\ref{R0}) in the form
\begin{equation}
	\frac{\delta_R \varepsilon(Z)}{B(E2)}=  
	\exp(\tilde{c}_0+\tilde{c}_1Z+\tilde{c}_2Z^2)Z^{\tilde{c}_3}\,\frac{\rm meV}{10^3{\rm fm}^5}\,, \label{eq:dRE2}
\end{equation}
where the coefficients $\tilde{c}_{0,1,2,3}$ are given in Table~\ref{tab:coefsE2} for $1s$, $2s$ and $2p_{1/2}$ states.

We point out that the results of this section readily generalize to superheavy elements. In tables \ref{Results_Ions_E1} and \ref{Results_Ions_E2} we present estimates of energy shifts for such elements up to $Z\leq 136$. 

\section{Effective potential for scalar nuclear polarization corrections in medium and heavy atoms}\label{Eff_Pot_Sec}

In Sec.~\ref{sec:heavy}, we calculated shifts of lowest energy levels in heavy hydrogen-like ions. This calculation may be performed with high accuracy because it makes use of exact electron wave functions in the discrete and continuous spectra. A generalization of these results to multielectron ions and neutral heavy atoms is hindered by many-body effects which are usually taken into account within the many-body theory based on the relativistic Hartree-Fock basis states. Precision calculation of energy level shifts in multielectron atoms and ions due to electric polarization of the nucleus goes beyond the scope of this paper, as it requires special numerical methods and computer codes which take into account many-body effects.

In this section, however, we will demonstrate that the effect of nuclear polarization may be taken into account by an effective potential which, when added to the unperturbed Hamiltonian, gives the same atomic energy level shifts as have been found in the previous section.\footnote{Rigorously  the energy shift may be presented as expectation value of a non-local (integration) self-energy operator $\Sigma(r,r',E)$ which at large distances becomes an ordinary local polarization potential  $V(r)=-\frac{\bar{\alpha}^{E1}_0 e^2}{2r^4}$. Such approach with the operator $\Sigma(r,r',E)$ added to the Hartree-Fock Hamiltonian  has been developed to calculate correlation corrections due to interaction between valence and core electrons \cite{Dzuba1987}.} Given that this potential is local and has a simple form, it may be added to the nuclear Coulomb potential and incorporated into  numerical calculations of the spectra of  multielectron ions and atoms including calculations of the isotope shifts. Such numerical computation will be given elsewhere. 

\subsection{General properties of the effective potential}

Recall that the effect of nuclear polarization due to the electron-nucleon interaction is well described by QED quantum corrections corresponding to Feynman graphs in Fig.~\ref{fig1}. In this process, the electron-nucleus interaction is essentially non-local as it is based on one-loop quantum effects with virtual electronic states having arbitrary high energy. At large distance, however, this interaction should reduce to a local four-point vertex. This dictates the large-distance asymptotic behaviour of the effective potential for this interaction,
$V_L(r)|_{r\to\infty} \sim r^{-2L-2}$, where $L=1$ for E1 and $L=2$ for E2 nuclear transitions, respectively. The coefficient of proportionality in this relation may be deduced from Eq.~(\ref{main-formula}). Indeed, at large distance from the nucleus, where the electron energy is small as compared with the energy of nuclear transitions, the energy shift should be proportional to the nuclear polarizability, $\Delta \varepsilon\propto \alpha^{EL}_0$, where 
\begin{equation}
    \alpha^{EL}_0 = \frac{8\pi}{2L+1} \frac{B(EL;L\to0)}{E_L}
    \label{nuc-pol}
\end{equation}
is the scalar nuclear polarizability due to the $EL$ nuclear transition with energy $E_L$. Therefore, we fix the asymptotic behaviour of the effective potential in the form 
\begin{equation}
V_L(r)|_{r\to\infty}  \to -\frac{e^2}2 \frac{\alpha^{EL}_0}{r^{2L+2}}\,.
\label{aassympt}
\end{equation}

Let $b$ be a characteristic distance at which the effective electron-nucleus interaction becomes non-local such that it cannot be described by the asymptotic formula (\ref{aassympt}). Although there may be different ways to extend the effective potential to the region $r<b$ which would have the same asymptotic behaviour (\ref{aassympt}),  we find it suitable to define the effective potential as
\begin{equation}
V_L(r)=-\frac{e^2}{2} \frac{\alpha^{EL}_0}{r^{2L+2}+b^{2L+2}}\,.
\label{eff-pot}
\end{equation}
In this case, the parameter $b$ may be thought of as an cut-off parameter below which the effective potential is nearly constant. 

We stress that $b$ is the only free parameter in the effective potential (\ref{eff-pot}). This parameter is, however, not universal in the sense that it should take into account specific nuclear properties such as nuclear charge $Z$, mass number $A$, nuclear radius $R_0$ and nuclear deformation $\beta_2$,
\begin{equation}
b=b(Z,A,R_0,\beta_2)\,. 
\label{b-function}
\end{equation}
Moreover, this parameter may be different for $s$ and $p_{1/2}$ states as well as for E1 and E2 nuclear transitions. Below, we describe the procedure that will allow us to find the dependence (\ref{b-function}) and will apply it to E1 and E2 nuclear transitions.

Let $\Delta\varepsilon^{(L)}_\Psi$ be an energy level shift in an atom or ion in a state $\Psi$ due to nuclear polarization induced by $EL$ nuclear transition. In particular, for hydrogen-like ions, the values of such shifts are calculated in the previous section and presented in Tables~\ref{Results_Ions_E1} and \ref{Results_Ions_E2}. Given the value of this energy level shift, we require that the effective potential (\ref{eff-pot}) should yield the same value,
\begin{equation}
    \Delta\varepsilon^{(L)}_\Psi = \langle \Psi | V_L |\Psi\rangle\,.
    \label{equation-for-b}
\end{equation}
This equation allows us to find the value of the free parameter $b$ in the effective potential (\ref{eff-pot}) for a given atom or ion in the state $\Psi$. The variety of values of this parameter for different $Z$, $A$, $R_0$ and $\beta_2$ sets up the function (\ref{b-function}). It is natural to expect that this function should vary in the range $R_0 < b \ll a_B/Z$.

It is important to note that the effective potential (\ref{eff-pot}) should be applied to multielectron atoms and ions. In this case, the parameter $b$ must be found from Eq.~(\ref{equation-for-b}) in which the energy shift in the left-hand side is calculated with a valence-electron wave function $\Psi$. In heavy atoms, this wave function (and the corresponding energy shift) may be quite different from the exact $1s$, $2s$ and $2p_{1/2}$ Dirac wave functions employed in Sect.~\ref{sec:heavy} for hydrogen-like ions. We will apply these functions only in the region $r\ll a_B/Z^{1/3}$ where the effects of the nuclear polarization are significant. In this region, calculation with the approximate radial wave functions (\ref{Bessel-functions}) would have a good accuracy. 

One has to keep in mind that the approximate wave functions (\ref{Bessel-functions}) correspond to the model of point-like nucleus. In heavy atoms, however, finite nuclear size corrections are significant. To take such effects into account in the leading order we will use the wave functions (\ref{Bessel-functions}) only outside the nucleus, i.e., for $R_0<r<a_B/Z^{1/3}$, while inside the nucleus, $0\leq r\leq R_0$, these functions may be extended as
\begin{equation}
	\begin{aligned}
		f_{s}(r)&=c_{ns}\frac{(-1+\gamma)J_{2\gamma}(x_0)- \frac{x_0}{2}J_{2\gamma-1}(x_0)}{R_0}\,,\\
		g_{s}(r)&=c_{ns}\frac{r}{R_0^2}Z\alpha J_{2\gamma}(x_0)\,,
	\end{aligned}\label{Bessel-inside}
\end{equation}
where $c_{ns}$ is the normalization constant and $x_0\equiv\sqrt{8ZR_0/a_B}$.

Note also that in atoms it is sufficient to consider only $s$-electron wave functions since higher waves give minor corrections due to nuclear polarization. Indeed, as is seen from Tables~\ref{Results_Ions_E1} and \ref{Results_Ions_E2}, the contribution from the $2p_{1/2}$ wave is from one to two orders in magnitude smaller than that from $2s$ wave. Therefore, we will restrict ourselves to specifying the effective potential for $s$ waves only.
From the comparison of the potentials  for $2p_{1/2}$ and $2s_{1/2}$ electrons we see that the difference between the $2p_{1/2}$ and $2s_{1/2}$ is not significant, so for the approximate calculation of a relatively small contribution of the  $p_{1/2}$ potential one may use the $s$-wave potential. 

\subsection{On the effective potential for light atoms}

Although in this paper we study nuclear polarization corrections to the spectra of medium and heavy atoms, in this section we briefly consider the effective potential in light atoms. We present this result only for a demonstration of the procedure of derivation of the effective potential which will be applied to medium and heavy atoms in subsequent subsections. In the case of light atoms, the corrections due to nuclear polarization were found analytically in Ref.~\cite{pachucki_nuclear-structure_1993}. Therefore, the procedure of constructing the effective potential is considerably simpler and more transparent in this case.

We restrict ourselves only to E1 nuclear transitions which are taken into account by the potential (\ref{eff-pot}) with $L=1$,
\begin{equation}
    V_1=-\frac{1}{2}\frac{e^2 \alpha_0^{E1} }{r^4+b^4}\,.\label{V1}
\end{equation}
In this potential, we have to determine the cut-off parameter $b$ as a function of nuclear parameters. 

Light atoms may be well-described by non-relativistic wave functions. Let $\phi(r)$ be a wave function of valence $s$ electron in a light atom. In this state, the expectation value of the operator (\ref{V1}) may be found analytically,
\begin{equation}
	\begin{aligned}
		\langle s| V_1|s \rangle
		&\approx-2\pi e^2 \bar{\alpha}_0^{E1}|\phi(0)|^2\int_0^{\infty}\frac{r^2 dr}{r^4+b^4}\\
		&=-\frac{\pi^2e^2 \bar{\alpha}_0^{E1}|\phi(0)|^2}{\sqrt{2}b}\,.\label{eq:s_integral}
	\end{aligned}
\end{equation}
Here we have taken into account that the $s$-wave function varies slowly inside the nucleus, so that $|\phi(r)|^2$ may be approximated by the electron density at the nucleus $|\phi(0)|^2$. The expectation value (\ref{eq:s_integral}) should be matched with the atomic energy $s$-level shift due to nuclear polarizability calculated in Ref.~\cite{pachucki_nuclear-structure_1993}:
\begin{equation}\label{eq:s_pachucki}
	\Delta\varepsilon=-m_e e^2
	|\phi(0)|^2\bar{\alpha}_0^{E1}\left[\frac{19}{6}+5 \ln\left(2\frac{\bar{\omega}}{m_e}\right)\right]\,.
\end{equation}
Here $\bar{\omega}$ is an average nuclear excitation energy in $E1$ transitions which varies from 5 MeV in deuterium to $\bar{\omega}$ given by Eq.~\eqref{baromega} in $^4$He and heavier nuclei. Equation $\Delta\varepsilon = \langle s| V_1|s \rangle$ yields the value of the cut-off parameter $b$ in light elements:
\begin{equation}\label{b_light}
	b\approx\frac{\pi^2}{\sqrt{2}\left[19/6+5\ln(2\bar{\omega}/m_e)\right]m_e}\,.
\end{equation}

Eq.~(\ref{b_light}) allows us to estimate the value of the cut-off parameter for light elements. In particular, for He and Li, $b\approx 105$ fm. As we will show below, in heavy atoms the value of the cut-off parameter is smaller but of the same order.

\subsection{Effective potential due to giant electric dipole resonance}\label{Atoms_E1}

In Sect.~\ref{Ions_E1}, we calculated shifts of $1s$, $2s$ and $2p_{1/2}$ energy levels in medium and heavy hydrogen-like ions. These data, however, do not apply to neutral heavy atoms in which energy level shifts come from $ns$ electronic orbitals with $n>2$. The energy level shifts in neutral atoms with $ns$ valence electrons will be denoted as $\Delta\varepsilon_{ns}$ in this section. In Table~\ref{E1_valence} we present the results of calculations of $\Delta\varepsilon_{ns}$ in medium and heavy atoms with $20\leq Z \leq 98$. This calculation is performed according to Eq.~(\ref{main-formula-E1}) making use of the $ns$ electron wave functions (\ref{Bessel-functions}) extended to the inside of the nucleus as in Eq.~(\ref{Bessel-inside}). Note that we do not specify the normalization coefficients $c_{ns}$ in these functions since our final result for the effective potential will be independent from these values.\footnote{Explicit values of these coefficients are presented, e.g., in Ref.~\cite{khriplovich_parity_1991}.} Therefore, in Table~\ref{E1_valence} we present the values of the dimensionless quantity $\Delta\varepsilon_{ns}/c_{ns}^2$.

In this section we consider the effects of nuclear polarization due to giant electric dipole resonance nuclear transition in medium and heavy atoms. In this case, the effective potential (\ref{eff-pot}) reads
\begin{equation}\label{Eff_Pot_E1}
	V_1=-\frac{1}{2}\frac{e^2 \alpha_0^{E1} }{r^4+b^4}\,,
	\quad
	\alpha_0^{E1}=\frac{ 8\pi B(E1)}{3E_{\rm GR}}\,,
\end{equation}
where $B(E1)$ is the reduced transition probability (\ref{nuc-strength}) and $E_{\rm GR}$ is the energy of giant electric dipole resonance transition (\ref{EGR}). The expectation value of the operator (\ref{Eff_Pot_E1}) in the $ns$ state reads
\begin{equation}
	\langle ns| V_1|ns\rangle =-\frac{1}{2}e^2\alpha_0^{E1}\int \frac{f_{ns}^2(r)+g_{ns}^2(r)}{r^4+b_{ns}^4}r^2dr\,,
	\label{V1ns}
\end{equation}
where $f_{ns}$ and $g_{ns}$ are radial wave functions (\ref{Bessel-functions}) extended to the inside of the nucleus as in Eqs.~(\ref{Bessel-inside}). 

Recall that the value of the parameter $b$ in the effective potential (\ref{Eff_Pot_E1}) (denoted by $b_{ns}$ in what follows) should be found from Eq.~(\ref{equation-for-b}). In the case of giant dipole resonance nuclear transitions, the energy in the left-hand side in Eq.~(\ref{equation-for-b}) is given by $\varepsilon_{ns}$ presented in Table~\ref{E1_valence} while the right-hand side is given by Eq.~(\ref{V1ns}). As a result, we have
\begin{equation}\label{Eqn_for_b_E1}
	\frac{1}{c_{ns}^2}\int \frac{f_{ns}^2(r)+g_{ns}^2(r)}{r^4+b_{ns}^4}r^2dr 
	= -\frac{3E_{\rm GR}}{4\pi e^2 B(E1)}\frac{\Delta\varepsilon_{ns}}{c_{ns}^2}\,.
\end{equation}
Eq.~(\ref{Eqn_for_b_E1}) defines the parameter $b_{ns}$ for each value of the energy shift $\Delta\varepsilon_{ns}/c_{ns}^2$. We solve this equation numerically and present the values of the parameter $b_{ns}$ in Table~\ref{E1_valence}. As is seen from this table, the parameter $b_{ns}$ is a monotonic function of $Z$, $b_{ns}=b_0(Z)$. It is convenient to approximate this function by the exponent, $b_0(Z)=\exp(\lambda_0 + \lambda_1 Z + \lambda_2 Z^2)Z^{\lambda_3}$, where the best fit for the parameters $\lambda_{0,1,2,3}$ is presented in Table~\ref{Coeff_Effective_Pot}.

The dependence of the parameter $b_{ns}$ on the mass number $A$ and nuclear radius $R$ may by taken into account in the same way as in Sect.~\ref{Ions_E1}: For each $Z$ we fix $A_0(Z)$ as in Eq.~(\ref{A0}) and consider $\Delta A$ deviations from $A_0(Z)$ (\ref{Delta_A_Def}). Then, for each isotope we fix the nuclear radius by Eq.~(\ref{R0}) and consider small deviation from this value, Eq.~(\ref{radius-deviation}). The parameter $b$ is now considered up to linear terms in $\Delta A$ and $\delta R$, $b_{ns}\equiv b(Z,A,R)=b_0(Z)+b_1(Z)\Delta A + b_2(Z)\delta R$. It is convenient to approximate the functions $b_{0,1,2}(Z)$ by exponents as follows:
\begin{equation}\label{Fit_b_E1}
	\begin{aligned}
		b(Z,A,R)&=\left[\exp(\lambda_0+\lambda_1Z+\lambda_2Z^2)Z^{\lambda_3}\right.\\
		        &\left.+\exp(\nu_0+\nu_1Z+\nu_2Z^2)Z^{\nu_3}\Delta A\right]\,{\rm fm} \\
		        &+\exp(\tau_0+\tau_1Z+\tau_2Z^2)Z^{\tau_3}\delta R\,.
	\end{aligned}
\end{equation}
Here $\lambda_{0,1,2,3}$, $\nu_{0,1,2,3}$ and $\tau_{0,1,2,3}$ are fitting parameters with numerical values given in Table~\ref{Coeff_Effective_Pot}. In this table, we present also the values for these coefficients which do the best fit of the parameters $b_{1s}$, $b_{2s}$ and $b_{2p_{1/2}}$ given in Table~\ref{E1_valence}. These values are given for comparison. In particular, it is seen that the functions $b_{ns}$ and $b_{2s}$ are very close, while the $1s$ and $2p_{1/2}$ state are described by slightly  different functions.
%Table VIII
\begin{center}
	\begin{table}[htb]
		\begin{tabular}{c|c|c}
			\hline\hline
			            & $s$                  & $p_{1/2}$              \\
			\hline
			$\lambda_0$	& $4.88$	           & $9.03$	            \\
			$\lambda_1$	& $-1.23\times10^{-2}$ & $1.32\times10^{-2}$	            \\
			$\lambda_2$	& $2.89\times10^{-5}$  & $4.05\times 10^{-5}$	\\
			$\lambda_3$	& $-0.178$  & $-1.46$	\\
			\hline
			$\nu_0$	    & $2.05$	           & $4.28$	                \\
			$\nu_1$	    & $-9.69\times10^{-3}$ & $-2.08\times10^{-2}$	            \\
			$\nu_2$	    & $2.02\times10^{-4}$  & $7.87\times10^{-5}$	\\
			$\nu_3$	    & $-1.07$	           & $-1.42$	                \\
			\hline
			$\tau_0$	& $1.94$	           & $10.6$	                \\
			$\tau_1$	& $-1.90\times10^{-2}$ & $3.99\times10^{-2}$	\\
			$\tau_2$	& $5.90\times10^{-5}$  & $-1.07\times10^{-4}$	\\
			$\tau_3$	& $-0.198$	           & $-2.77$	            \\
			\hline\hline
		\end{tabular}
		\caption{Numerical values of the fitting coefficients in Eq.~\eqref{Fit_b_E1} describing the cut-off parameter $b$ for $s$ and $p_{1/2}$ effective potentials in multielectron atoms due to nuclear E1 electric dipole giant resonance.}\label{Coeff_Effective_Pot}
	\end{table}	
\end{center}

To summarize, the function (\ref{Fit_b_E1}) with the parameters $\lambda_{0,1,2,3}$, $\nu_{0,1,2,3}$ and $\tau_{0,1,2,3}$ given in Table~\ref{Coeff_Effective_Pot} specifies the effective potential (\ref{Eff_Pot_E1}). This effective potential allows one to calculate the energy level shifts with error under 1\% as compared with the results presented in Table \ref{E1_valence}.

\subsection{Effective potential due to electric quadrupole nuclear polarization}

In the derivation of the effective potential which takes into account electric quadrupole nuclear polarization corrections in the atomic spectra we will follow the same procedure as in Sect.~\ref{Atoms_E1}. Namely, we start with the calculation of the energy shifts $\Delta\tilde\varepsilon_{ns}$ using the $ns$ valence electron wave functions (\ref{Bessel-functions}) as the initial and final electronic states in Eq.~(\ref{main-formula-E2}). The results of these calculations are presented in Table~\ref{E2_valence}. 

In the case of nuclear polarization due to the rotational nuclear transitions the effective potential (\ref{eff-pot}) reads
\begin{equation}\label{Eff_Pot_E2}
V_2=-\frac12\frac{e^2\bar{\alpha}_0^{E2}}{r^6+\tilde{b}^6}\,,
\end{equation}
where $\tilde{b}$ is the effective cut-off parameter and
\begin{equation}\label{alpha_0_E2}
	\bar{\alpha}_0^{E2}=\frac{8\pi}{5}\frac{B(E2)}{\bar E_{\rm rot}}\,,
	\quad \bar E_{\rm rot}=50 \ {\rm keV},
\end{equation}
is the modified E2 nuclear polarizability. In contrast with the conventional nuclear polarizability (\ref{nuc-pol}), it has fixed energy $\bar E_{\rm rot}=50$ keV which corresponds to typical energy of the lowest rotational state in deformed heavy nuclei. Using this fixed energy in Eq.~(\ref{alpha_0_E2}) appears more convenient in the effective potential (\ref{Eff_Pot_E2}) because the physical energy (\ref{rot-energy}) depends on the deformation parameter $\beta_2$ which changes non-monotonically with $Z$. Indeed, the potential  (\ref{Eff_Pot_E2}) defined via the modified nuclear polarizability (\ref{alpha_0_E2}) is a monotonic function of $Z$ and, thus, it is suitable for modelling atomic energy level shifts due to the nuclear rotational transitions given in Table~\ref{E2_valence}.

The expectation value of the effective potential in the $ns$ state is
\begin{equation}\label{Expect_VP_E2}
	\langle ns| V_2| ns\rangle=-\frac{4\pi B(E2)e^2}{5\bar{E}_{\rm rot}}
	\int \frac{f_{ns}^2+g_{ns}^2}{r^6+\tilde{b}^6}r^2dr\,,
\end{equation}
where the integral is calculated with the use of the wave functions \eqref{Bessel-functions} extended to the inside of the nucleus as in Eq.~\eqref{Bessel-inside}. The cut-off parameter $\tilde b$ should now be found upon matching the expectation value of the effective operator (\ref{Expect_VP_E2}) with the energy shifts $\Delta\tilde\varepsilon_{ns}$ given in Table~\ref{E2_valence},
\begin{equation}
	\frac{1}{c_{ns}^2}\int \frac{f_{ns}^2+g_{ns}^2}{r^6+\tilde{b}^6}r^2dr 
	=-\frac{5\bar{E}_{\rm rot}}{4\pi e^2}\frac{\Delta\tilde\varepsilon_{ns}}{c_{ns}^2B(E2)}\,.
\end{equation}
The value of the parameter $\tilde b$ may be found by solving this equation numerically for each given isotope. We present these values in Table~\ref{E2_valence}. In the same table we give also the values of this parameter corresponding to the $1s$, $2s$ and $2p_{1/2}$ energy level shifts in hydrogen-like ions from Table~\ref{Results_Ions_E2}. We point out that the values of the parameter $b_{ns}$ are very close to the corresponding values of $b_{1s}$ and especially $b_{2s}$ that confirms the consistency of the definition of the effective potential (\ref{Eff_Pot_E2}).

The numerical values of the parameter $\tilde b_{ns}$ in Table~\ref{E2_valence} define the function $\tilde b_{ns}=\tilde b(Z,A,R)$. This function may be approximated by
\begin{equation}\label{fit_b_tilde_E2}
	\begin{aligned}
		\tilde{b}(Z,A,R)&=\big[\tilde{\lambda}_0+\tilde{\lambda}_1Z\\
		&+\exp(\tilde{\nu}_0+\tilde{\nu}_1Z+\tilde{\nu}_2Z^2)\Delta A\big] \,{\rm fm} \\
		&+\exp(\tilde{\tau}_0+\tilde{\tau}_1Z+\tilde{\tau}_2Z^2)Z^{\tau_3}\delta R\,,
	\end{aligned}
\end{equation}
where the best fit for the parameters $\tilde\lambda_{0,1,2}$, $\tilde \nu_{0,1,2}$ and $\tilde\tau_{0,1,2}$ is given in Table~\ref{Coeff_Effective_Pot_E2}.
%Table X
\begin{center}
	\begin{table}[htb]
		\begin{tabular}{c|c|c}
			\hline\hline
			                    & $s$                  & $p_{1/2}$               \\
			\hline
			$\tilde{\lambda}_0$ & $135$	               & $148$                   \\
			$\tilde{\lambda}_1$	& $-0.600$	           & $-0.682$	             \\
			\hline
			$\tilde{\nu}_0$	    & $-0.172$	           & $7.27\times10^{-2}$	 \\
			$\tilde{\nu}_1$	    & $-3.51\times10^{-2}$ & $-3.59\times10^{-2}$	 \\
			$\tilde{\nu}_2$	    & $9.47\times10^{-5}$  & $8.79 \times 10^{-5}$	 \\
			\hline
			$\tilde{\tau}_0$	& $1.93$	           &	$2.28$               \\
			$\tilde{\tau}_1$	& $-2.36\times10^{-2}$ &	$-2.33\times10^{-2}$ \\
			$\tilde{\tau}_2$	& $5.06\times10^{-5}$  &	$4.54\times10^{-5}$	 \\
			$\tilde{\tau}_3$	& $0.386$              &	$0.344$	             \\
			\hline\hline
		\end{tabular}
		\caption{Numerical values of the coefficients $\tilde\lambda_{0,1}$, $\tilde\nu_{0,1,2}$ and $\tilde\tau_{0,1,2,3}$ which specify the cut-off parameter (\ref{fit_b_tilde_E2}) as a function of $Z$, $A$ and $R$.}\label{Coeff_Effective_Pot_E2}
	\end{table}	
\end{center}

To summarize, in this section we found the effective potential (\ref{Eff_Pot_E2}) with the parameter $\tilde b$ given by Eq.~(\ref{fit_b_tilde_E2}). For medium and heavy elements, this effective potential reproduces the atomic energy level shifts presented in Table \ref{E2_valence} with error under 1\%. For superheavy elements, the error is within 5\%.

\section{Conclusions}\label{Conclusion}
	
In this paper, we studied the effects of electric nuclear polarization in the spectra of medium and heavy atoms and ions. These effects manifest themselves in electron energy levels shifts and isotope shifts  or in contributions to the hyperfine structure. Although in neutral atoms such effects are small, they are strongly enhanced in heavy hydrogen-like ions. Indeed, the $s_{1/2}$ and $p_{1/2}$ electron wave functions are known to be significantly enhanced near a heavy nucleus \cite{khriplovich_parity_1991}, so that corrections due to nuclear polarization are observable.
	
Recall that the tensor nuclear polarizability is responsible for contributions to the hyperfine structure. We observe that the effective operator describing the contributions from the tensor polarizability has the same tensor structure as the quadrupole nuclear moment. Therefore, it is natural to compare the contributions from these operators to the atomic hyperfine structure. We show that the effect from the tensor nuclear polarizability is nearly three orders of magnitude weaker than that from the electric quadrupole nuclear moment. Although in neutral atoms this effect is rather unobservable, it may be noticeable in heavy hydrogen-like ions where the hyperfine energy splitting is on order of 1 eV. In this paper, we estimated the order of magnitude of this effect while accurate calculations of contributions from the tensor polarizability in particular atoms are left for further studies.

The scalar nuclear polarizability is responsible for atomic energy levels shifts. The method of calculating these shifts was developed in a series of papers \cite{plunien_nuclear_1989,plunien_nuclear_1991,plunien_nuclear-polarization_1995,plunien_erratum:_1996,nefiodov_nuclear_1996}
where these shifts were found for a limited number of hydrogen-like ions. We employ this method and extend the results to include hydrogen-like ions with $20\leq Z\leq 98$ and find an interpolating formula which reproduces these shifts as a function of $Z$, $A$ and nuclear radius. 
Errors in the atomic calculations are smaller than 5\%.

Energy level shifts in some superheavy hydrogen-like ions ($Z=106, 114, 122, 130, 136$) are also calculated. The results for other superheavy elements may be estimated using a two-point interpolation formula based on the considered ions. The energy shift grows with $Z$ approximately exponentially. Therefore,  this functions may be taken in the form $\varepsilon(Z)=\varepsilon_0 e^{a Z}$, with  parameters $\varepsilon_0$ and $a$ chosen to reproduce exactly values of $\varepsilon(Z)$ for two nearby ions where we have performed the calculations.  

We consider separately contributions from nuclear giant electric dipole resonance transitions (E1) and rotational transitions (E2). In the latter case, the energy shifts depend also on the nuclear quadrupole deformation parameter $\beta_2$. By comparing our results with the earlier calculations in some heavy hydrogen-like ions  \cite{plunien_nuclear_1991,plunien_nuclear-polarization_1995} we find that the obtained formulae provide energy shifts in heavy elements with error of atomic calculations under 7\%. The accuracy in the nuclear parameters, which we use, is defined in the referenced papers. 

We also study the dependence of the energy shifts on the variation of nuclear radius $\Delta R$. Indeed, certain isotopes may have deviations of the nuclear radius from the general rule (\ref{R0}), and the energy shifts are sensitive to such deviations. This may lead to nonlinearity in King's plot for isotope shifts and imitate effects of new interactions. Note that such nonlinearity was observed recently in Yb isotopes \cite{counts_3sigma_2020} and interpreted in terms of new interaction  beyond the Standard Model. Our calculation opens the way for systematic study of this effect in a wide range of atoms and ions.

We point out that the calculation of corrections due to nuclear polarizability in the spectra of hydrogen-like ions may be performed with a good accuracy because in this calculation one can use exact Dirac wave function in discrete and continuous spectra. A generalization of these results to multielectron ions and neutral atoms is hindered by many-body effects which are usually taken into account using many-body theory based on the relativistic  Hartree-Fock basis  states. In order to facilitate such calculations in future works, in this paper we develop an effective potential which models the corrections due to the nuclear polarizability. This potential has simple local form (\ref{eff-pot}) with one parameter $b$ which is approximated by the functions (\ref{Fit_b_E1}) and (\ref{fit_b_tilde_E2}) such that the expectation value of this potential gives correct energy level shifts for hydrogen-like ions and for $s$ electrons in a many-electron atoms. As a result, to calculate the nuclear polarization effect in many-electron atoms and ions, one simply has to add this potential to the nuclear Coulomb interaction and then solve for the self-consistent Hartree-Fock equations. Incorporation into the calculation of the correlation corrections is straightforward with the use of Hartree-Fock basis states.

The effect is dominated by the nuclear polarization potential for $s$ electrons, with a much smaller contribution from the slightly different  potential for $p_{1/2}$ electrons. The difference between 2$s$ and $ns$ potentials is very small, therefore, one can use $2p_{1/2}$ potential for all $p_{1/2}$ electrons. Moreover, the difference between the $p_{1/2}$ and $s$ potentials is not significant. Potentials in all waves have the same long distance asymptotic but have different cut-off parameters $b$. We have checked that in heavy atoms the difference between parameters $b$ in 2$s$ and 2$p_{1/2}$ potentials is small. Also, there is no significant dependence on the principal quantum number $n$, parameters $b$ for 1$s$, 2$s$ and high $ns$ electrons are practically the same.  For simplicity, one may use the $ns$-wave potential for all waves. Indeed, direct contributions of this potential in $p_{3/2}$ and  higher waves may be neglected since they do not come close to the nucleus, but the shift of their energies appears due to the so called core polarization effect: the change of the  $s$-wave electron wave functions leads to the change of the  self-consistent Hartree-Fock potential affecting all electrons. Such calculations will be done elsewhere.

\subsection*{Acknowledgements}
This work was supported by the Australian Research Council Grants No. DP190100974 and DP200100150 and the Gutenberg Fellowship. The authors are grateful to Savely Karshenboim, Vladimir Shabaev, Antonios Karantzias, Amy Geddes and Julian Berengut for useful discussions.

%%%%%%%%%%%%%%%%%%%%%%%%%%%%%%%%%%%%%%%%%%%%%%%%%%%%%%%%%%%%%%%%%%%%%

%

%%%%%%%%%%%%%%%%%%%%%%%%%%%%%%%%%%%%%%%%%%%%%%%%%%%%%%%%%%%%%%%%%%%%%
%Table I
\begin{center}
		\begin{table*}[tbh]
			\begin{tabular}{cc| c | c| c|c|c|c}
				\hline\hline
				\multirow{2}{*}{$Z$}	&\multirow{2}{*}{$A$}	&$-\Delta \varepsilon_{1s}$	&$-\Delta \varepsilon_{2s}$		&$-\Delta \varepsilon_{2p_{1/2}}$ &$\delta_R\varepsilon_{1s}$	&$\delta_R\varepsilon_{2s}$		&$\delta_R\varepsilon_{2p_{1/2}}$\\
				&&(meV)&(meV)&(meV)&(meV/fm)&(meV/fm)&(meV/fm)\\
				\hline
				20  &   40  &   $3.93\times 10^{-3}$ &  $5.01\times 10^{-4}$ & $1.84\times 10^{-6}$ & \multirow{2}{*}{$3.53\times 10^{-4}$} & \multirow{2}{*}{$4.48\times 10^{-5}$} & \multirow{2}{*}{$1.69\times10^{-7}$} \\
				20  &   44  &   $4.37\times 10^{-3}$ &  $5.57\times 10^{-4}$ & $2.04\times 10^{-6}$ & & & \\ 
				\hline
				26	&	54	&	$	1.53 \times 10^{-2}	$	&	$	1.97 \times 10^{-3}	$	&	$	1.24 \times 10^{-5}	$ & \multirow{2}{*}{$1.45\times 10^{-3}$} & \multirow{2}{*}{$1.88\times 10^{-4}$} & \multirow{2}{*}{$1.22\times10^{-6}$} 	\\
				26	&	60	&	$	1.70 \times 10^{-2}	$	&	$	2.20 \times 10^{-3}	$	&	$	1.39 \times 10^{-5}	$ & & &	\\
				\hline
				34	&	72	&	$	6.29 \times 10^{-2}	$	&	$	8.33 \times 10^{-3}	$	&	$	9.23 \times 10^{-5}	$ & \multirow{2}{*}{$6.53\times 10^{-3}$} & \multirow{2}{*}{$8.64\times 10^{-4}$} & \multirow{2}{*}{$9.80\times10^{-6}$} 		\\
				34	&	80	&	$	7.00 \times 10^{-2}	$	&	$	9.26 \times 10^{-3}	$	&	$	1.03 \times 10^{-4}	$ & & &		\\
				\hline
				42	&	92	&	$	0.207	$	&	$	2.83 \times 10^{-2}	$	&	$	4.93 \times 10^{-4}	$ & \multirow{2}{*}{$2.08\times 10^{-2}$} & \multirow{2}{*}{$2.84\times 10^{-3}$} & \multirow{2}{*}{$5.03\times10^{-5}$} 			\\
				42	&	100	&	$	0.225	$	&	$	3.07 \times 10^{-2}	$	&	$	5.36 \times 10^{-4}	$ & & &	\\
				\hline
				50	&	112	&	$	0.583	$	&	$	8.23 \times 10^{-2}	$	&	$	2.11 \times 10^{-3}	$ & \multirow{2}{*}{$6.44\times 10^{-2}$} & \multirow{2}{*}{$9.09\times 10^{-3}$} & \multirow{2}{*}{$2.36\times10^{-4}$} 				\\
				50	&	126	&	$	0.649	$	&	$	9.17 \times 10^{-2}	$	&	$	2.35 \times 10^{-3}	$ & & &	\\
				\hline
				58	&	136	&	$	1.51	$	&	$	0.223	$	&	$	8.00 \times 10^{-3}	$ & \multirow{2}{*}{$0.167$} & \multirow{2}{*}{$2.46\times10^{-2}$} & \multirow{2}{*}{$8.91\times10^{-4}$} 					\\
				58	&	142	&	$	1.57	$	&	$	0.232	$	&	$	8.31 \times 10^{-3}	$ & & &	\\
				\hline
				66	&	154	&	$	3.51	$	&	$	0.543	$	&	$	2.64 \times 10^{-2}	$ & \multirow{2}{*}{$0.409$} & \multirow{2}{*}{$6.32\times10^{-2}$} & \multirow{2}{*}{$3.09\times10^{-3}$} 					\\
				66	&	164	&	$	3.70	$	&	$	0.573	$	&	$	2.79 \times 10^{-2}	$ & & &	\\
				\hline
				74	&	180	&	$	8.11	$	&	$	1.33	$	&	$	8.55 \times 10^{-2}	$ & \multirow{2}{*}{$0.974$} & \multirow{2}{*}{$0.159$} & \multirow{2}{*}{$1.02\times10^{-2}$} 					\\
				74	&	186	&	$	8.33	$	&	$	1.36	$	&	$	8.78 \times 10^{-2}	$ & & &	\\
				\hline
				82	&	204	&	$	18.0	$	&	$	3.13	$	&	$	0.262	$ & \multirow{2}{*}{$2.22$} & \multirow{2}{*}{$0.388$} & \multirow{2}{*}{$3.22\times10^{-2}$} 					\\
				82	&	210	&	$	18.4	$	&	$	3.20	$	&	$	0.268	$ & & &	\\
				\hline
				90	&	228	&	$	39.1	$	&	$	7.33	$	&	$	0.792	$ & \multirow{2}{*}{$5.13$} & \multirow{2}{*}{$0.963$} & \multirow{2}{*}{$0.103$} 					\\
				90	&	230	&	$	39.4	$	&	$	7.38	$	&	$	0.797	$ & & &	\\
				\hline
				92	&	234	&	$	47.5	$	&	$	9.07	$	&	$	1.04	$ & \multirow{2}{*}{$6.34$} & \multirow{2}{*}{$1.21$} & \multirow{2}{*}{$0.137$} 					\\
				92	&	240	&	$	48.4	$	&	$	9.23	$	&	$	1.06	$ & & &	\\
				\hline
				98	&	248	&	$	84.3	$	&	$	17.1	$	&	$	2.38	$ & \multirow{2}{*}{$12.2$} & \multirow{2}{*}{$2.48$} & \multirow{2}{*}{$0.339$} 					\\
				98	&	252	&	$	85.3	$	&	$	17.3	$	&	$	2.40	$ & & &	\\
				\hline
				106	&	272	&	$	186	$	&	$	41.2	$	&	$	7.39	$ & \multirow{2}{*}{$29.0$} & \multirow{2}{*}{$6.44$} & \multirow{2}{*}{$1.13$} 					\\	
				106	&	274	&	$	187	$	&	$	41.4	$	&	$	7.42	$ & & &	\\
				\hline
				114	&	294	&	$	426	$	&	$	104	$	&	$	24.4	$ & \multirow{2}{*}{$71.3$} & \multirow{2}{*}{$17.5$} & \multirow{2}{*}{$4.10$} 					\\	
				114	&	296	&	$	427	$	&	$	104	$	&	$	24.5	$ & & &	\\
				\hline
				122	&	316	&	$	1050	$	&	$  286	$	&	$	90.8	$ & \multirow{2}{*}{$195$} & \multirow{2}{*}{$53.6$} & \multirow{2}{*}{$16.5$} 					\\	
				122	&	318	&	$	1053	$	&	$	287	$	&	$	91.1	$ & & &	\\
				\hline
				130	&	338	&	$	3046	$	&	$  936	$	&	$	437  	$ & \multirow{2}{*}{$642$} & \multirow{2}{*}{$200$} & \multirow{2}{*}{$90.5$} 					\\	
				130&	340	&	$	3054	$	&	$  939	$	&	$	438  	$ & & &	\\
				\hline
				136	&	354	&	$	9972	$	&	$  3334	$	&	$	2629	$ & \multirow{2}{*}{$2427$} & \multirow{2}{*}{$829$} & \multirow{2}{*}{$629$} 					\\	
				136	&	356	&	$	9987	$	&	$  3339	$	&	$	2633	$ & & &	\\		\hline\hline
			\end{tabular}
			\caption{Nuclear giant electric dipole resonance contributions to the energy shifts of the $1s$, $2s$ and $2p_{1/2}$ levels in hydrogen-like ions due to nuclear polarization effects. The values of the coefficient $\delta_R\varepsilon$ characterizing the linear dependence of $\Delta\varepsilon$ on the nuclear radius variation $\Delta R$ are also presented.}\label{Results_Ions_E1}
\end{table*}
	
%Table IV
\begin{table*}[htb]
			\begin{tabular}{cc| c | c| c|c|c|c}
				\hline\hline
				\multirow{2}{*}{$Z$}	&	\multirow{2}{*}{$A$}	&	$-\frac{\Delta \varepsilon_{1s}}{B(E2)}$&	$-\frac{\Delta \varepsilon_{2s}}{B(E2)}$	&	$-\frac{\Delta \varepsilon_{2p{1/2}}}{B(E2)}$ &	$-\frac{\delta_R\varepsilon_{1s}}{B(E2)}$&	$-\frac{\delta_R\varepsilon_{2s}}{B(E2)}$	&	$-\frac{\delta_R\varepsilon_{2p{1/2}}}{B(E2)}$		\\
				&&$\left(\frac{{\rm meV}}{10^{3}{\rm fm}^4}\right)$&$\left(\frac{{\rm meV}}{10^{3}{\rm fm}^4}\right)$&$\left(\frac{{\rm meV}}{10^{3}{\rm fm}^4}\right)$ &$\left(\frac{{\rm meV}}{10^{3}{\rm fm}^5}\right)$&$\left(\frac{{\rm meV}}{10^{3}{\rm fm}^5}\right)$&$\left(\frac{{\rm meV}}{10^{3}{\rm fm}^5}\right)$\\
				\hline
				26	&	54	&	$	2.00	$	&	$	0.259	$	&	$	1.59 \times 10^{-3}	$ & \multirow{2}{*}{$0.680$}	 & \multirow{2}{*}{$8.80\times 10^{-2}$} & \multirow{2}{*}{$5.70\times10^{-4}$} \\
				26	&	56	&	$	1.96	$	&	$	0.253	$	&	$	1.56 \times 10^{-3}	$ & & &	\\
				\hline
				34	&	74	&	$	5.44	$	&	$	0.721	$	&	$	7.88 \times 10^{-3}	$ & \multirow{2}{*}{$1.86$}	 & \multirow{2}{*}{$0.245$} & \multirow{2}{*}{$2.78\times10^{-3}$} 	\\
				34	&	76	&	$	5.37	$	&	$	0.711	$	&	$	7.76 \times 10^{-3}	$ & & &	\\
				\hline
				44	&	102	&	$	14.6	$	&	$	2.01	$	&	$	3.83 \times 10^{-2}	$ & \multirow{2}{*}{$4.88$}	 & \multirow{2}{*}{$0.671$} & \multirow{2}{*}{$1.32\times10^{-2}$} 	\\
				44	&	104	&	$	14.4	$	&	$	1.98	$	&	$	3.77 \times 10^{-2}	$ & & &	\\
				\hline
				46	&	108	&	$	17.4	$	&	$	2.41	$	&	$	5.08 \times 10^{-2}	$ & \multirow{2}{*}{$5.48$}	 & \multirow{2}{*}{$0.761$} & \multirow{2}{*}{$1.64\times10^{-2}$} 	\\
				46	&	110	&	$	17.2	$	&	$	2.38	$	&	$	5.01 \times 10^{-2}	$ & & &	\\
				\hline
				62	&	152	&	$	68.3	$	&	$	10.3 $	&	$	0.43	$ & \multirow{2}{*}{$22.1$}	 & \multirow{2}{*}{$3.34$} & \multirow{2}{*}{$0.141$} 	\\
				62	&	154	&	$	67.7	$	&	$	10.2 $	&	$	0.42	$ & & &	\\
				\hline
				66	&	162	&	$	91.2	$	&	$	14.6 $	&	$	0.71	$ & \multirow{2}{*}{$30.9$}	 & \multirow{2}{*}{$4.78$} & \multirow{2}{*}{$0.233$} 	\\
				66	&	164	&	$	93.3	$	&	$	14.5 $	&	$	0.70	$ & & &	\\
				\hline
				70	&	172	&	$	129	    $	&	$	20.6 $	&	$	1.14	$ & \multirow{2}{*}{$41.0$}	 & \multirow{2}{*}{$6.54$} & \multirow{2}{*}{$0.368$} 	\\
				70	&	174	&	$	128	    $	&	$	20.4 $	&	$	1.13	$ & & &	\\
				\hline
				74	&	184	&	$	175	    $   &	$	28.7 $	&	$	1.83	$ & \multirow{2}{*}{$53.5$}	 & \multirow{2}{*}{$8.79$} & \multirow{2}{*}{$0.565$} 	\\
				74	&	186	&	$	173	    $	&	$	28.5 $	&	$	1.81	$ & & &	\\
				\hline
				90	&	228	&	$	607	    $	&	$	114	$	&	$	12.3	$ & \multirow{2}{*}{$185$}	 & \multirow{2}{*}{$34.9$} & \multirow{2}{*}{$3.74$} 	\\
				90	&	230	&	$	602	    $	&	$	114	$	&	$	12.2	$ & & &	\\
				\hline
				92	&	234	&	$	710     $	&	$	136	$	&	$	15.6	$ & \multirow{2}{*}{$220$}	 & \multirow{2}{*}{$42.2$} & \multirow{2}{*}{$4.81$} 	\\
				92	&	236	&	$	705	    $	&	$	135	$	&	$	15.5	$ & & &	\\
				\hline
				98	&	250	&	$	1151	$   &	$	235	$   &	$	32.4	$ & \multirow{2}{*}{$360$}	 & \multirow{2}{*}{$73.4$} & \multirow{2}{*}{$10.1$} 	\\
				98	&	252	&	$	1143	$	&	$	234	$	&	$	32.2	$ &  & &	\\
				\hline
				106	&	272	&	$	2249	$   &	$	503	$   &	$	89.6	$ & \multirow{2}{*}{$ 706 $}	 & \multirow{2}{*}{$ 158 $} & \multirow{2}{*}{$ 28.0 $} 	\\
				106	&	274	&	$	2234	$	&	$	500	$	&	$	89.0	$ &  & &	\\
				\hline
				114	&	294	&	$	2249	$   &	$	1147	$   &	$	268	$ & \multirow{2}{*}{$ 1503$}	 & \multirow{2}{*}{$ 372 $} & \multirow{2}{*}{$ 86.0 $} 	\\
				114	&	296	&	$	2234	$	&	$	1140	$	&	$	266	$ &  & &	\\ 
				\hline		
				122	&	316	&	$	1.04\times 10^4	$   &	$	2885	$   &	$	913	$ & \multirow{2}{*}{$ 3639 $}	 & \multirow{2}{*}{$ 1009 $} & \multirow{2}{*}{$ 315 $} 	\\
				122	&	318	&	$	1.03\times 10^4	$	&	$	2861	$	&	$	907	$ &  & &\\		\hline
				130	&	338	&	$	2.79\times 10^4	$   &	$	8761	$   &	$	4090	$ & \multirow{2}{*}{$1.06\times 10^4 $}	 & \multirow{2}{*}{$ 3328 $} & \multirow{2}{*}{$ 1526 $} 	\\
				130	&	340	&	$	2.77\times 10^4	$	&	$	8701	$	&	$	4062	$ &  & &\\	\hline
				136	&	354	&	$	8.80\times 10^4	$   &	$	3.05\times 10^4	$   &	$	2.42\times 10^4	$ & \multirow{2}{*}{$ 3.64\times 10^4 $}	 & \multirow{2}{*}{$ 1.26\times 10^3 $} & \multirow{2}{*}{$ 9756 $} 	\\
				136	&	356	&	$	8.74\times 10^4	$	&	$	3.03\times 10^4	$	&	$	2.40\times 10^4	$ &  & &	\\	\hline\hline
			\end{tabular}
			\caption{Nuclear rotational transition contributions to the energy shifts of the $1s$, $2s$ and $2p_{1/2}$ levels in hydrogen-like ions due to nuclear polarization effects. The numbers in the table need to be multiplied by the reduced nuclear transition probability $B(E2)\equiv B(E2;2\rightarrow0)$ to give the actual energy shifts. The values of $B(E2)$ for different nuclei may be found, e.g., in Ref.~\cite{Raman:1201zz}. The values of the coefficient $\delta_R\varepsilon/B(E2)$ characterizing the linear dependence of $\Delta\varepsilon/B(E2)$ on the nuclear radius variation $\Delta R$ are also presented.}\label{Results_Ions_E2}
\end{table*}

%Table VII

\begin{table*}[htb]
		\begin{tabular}{cc|c|c|c|c|c|c|c}
			\hline\hline
			\multirow{2}{*}{$Z$} & \multirow{2}{*}{$A$} & \multirow{2}{*}{$-\frac{\Delta \varepsilon_{ns}/c_{ns}^2}{10^{-7}}$} & $b_{ns}$ & $b_{1s}$ & $b_{2s}$ & $b_{2p_{1/2}}$ & \multirow{2}{*}{$\delta_Rb_s$} &  \multirow{2}{*}{$\delta_Rb_{p_{1/2}}$} \\
			&&& (${\rm fm}$) & (${\rm fm}$) & (${\rm fm}$) & (${\rm fm}$) & & \\
			\hline
			20	&	40	&	$5.35\times 10^{-4}$	& 62.0 & 62.3 & 62.1 & 141 & \multirow{2}{*}{$4.91$} & \multirow{2}{*}{$22.8$} 	\\
			20	&	44	&	$5.95 \times 10^{-4}$	& 62.9 & 63.3 & 63.1 & 145 & &  \\
			\hline
			26	&	54	&	$1.62 \times 10^{-3}$	& 54.9 & 55.3 & 55.0 & 94.8 & \multirow{2}{*}{$4.52$} &	\multirow{2}{*}{$12.4$}\\
			26	&	60	&	$1.81 \times 10^{-3}$	& 56.0 & 56.4 & 59.5 & 97.7 & &	\\
			\hline
			34	&	74	&	$5.34 \times 10^{-3}$	& 47.9 & 48.4 & 48.1 & 70.2 & \multirow{2}{*}{$4.02$} & \multirow{2}{*}{$8.10$} \\
			34	&	76	&	$5.49 \times 10^{-3}$	& 48.2 & 48.7 & 48.3 & 70.7 & &	\\
			\hline
			42	&	96	&	$1.48 \times 10^{-2}$	& 42.6 & 43.1 & 42.7 & 57.0 & \multirow{2}{*}{$3.32$} & \multirow{2}{*}{$5.58$} \\
			42	&	98	&	$1.51 \times 10^{-2}$	& 42.8 & 43.3 & 42.9 & 57.4 & &	\\
			\hline
			50	&	118	&	$3.61 \times 10^{-2}$	& 38.2 & 38.7 & 38.3 & 48.2 & \multirow{2}{*}{$3.15$} & \multirow{2}{*}{$4.75$}	\\
			50	&	120	&	$3.66 \times 10^{-2}$	& 38.4 & 38.9 & 38.5 & 48.5 & &	\\
			\hline
			58	&	140	&	$8.17 \times 10^{-2}$	& 34.6 & 35.1 & 34.7 & 41.8 & \multirow{2}{*}{$2.83$} & \multirow{2}{*}{$3.94$} \\
			58	&	142	&	$8.22 \times 10^{-2}$	& 34.7 & 35.3 & 34.8 & 42.0 &  &	\\
			\hline
			66	&	162	&	$0.174$	                & 31.5 & 32.0 & 31.6 & 36.9 & \multirow{2}{*}{$2.55$} & \multirow{2}{*}{$3.35$} \\
			66	&	164	&	$0.176$                 & 31.6 & 32.1 & 31.7 & 37.0 & & \\
			\hline
			74	&	184	&	$0.363$	                & 28.9 & 29.4 & 29.0 & 33.0 & \multirow{2}{*}{$2.37$} & \multirow{2}{*}{$2.98$}	\\
			74	&	186	&	$0.366$	                & 29.0 & 29.5 & 29.1 & 33.1 & &	\\
			\hline
			82	&	206	&	$0.746$	                & 26.7 & 27.2 & 26.8 & 29.8 & \multirow{2}{*}{$2.17$} & \multirow{2}{*}{$2.62$}	\\
			82	&	208	&	$0.752$	                & 26.7 & 27.3 & 26.9 & 29.9 & &	\\
			\hline
			90	&	228	&	$1.53$	                & 24.7 & 25.3 & 24.9 & 27.2 & \multirow{2}{*}{$2.04$} & \multirow{2}{*}{$2.38$}	\\
			90	&	230	&	$1.54$	                & 24.8 & 25.3 & 24.9 & 27.2 & &	\\
			\hline
			98	&	250	&	$3.17$	                & 23.1 & 23.6 & 23.2 & 25.0 & \multirow{2}{*}{$1.96$} & \multirow{2}{*}{$2.24$}	\\
			98	&	252	&	$3.19$	                & 23.1 & 23.6 & 23.3 & 25.0 & &	\\
			\hline
			106	&	272	&	$6.69$	                & 21.7 & 22.2 & 21.8 & 23.2 & \multirow{2}{*}{$1.87$} & \multirow{2}{*}{$2.08$}	\\
			106	&	274	&	$6.72$	                & 21.7 & 22.2 & 21.9 & 23.3 & &	\\
			\hline
			114	&	294	&	$14.7$	                & 20.4 & 20.9 & 20.6 & 21.6 & \multirow{2}{*}{$1.75$} & \multirow{2}{*}{$1.91$}	\\
			114	&	296	&	$14.7$	                & 20.5 & 21.0 & 20.6 & 21.7 & &	\\
			\hline
			122	&	316	&	$34.1$	                & 19.4 & 19.8 & 19.5 & 20.3 & \multirow{2}{*}{$1.68$} & \multirow{2}{*}{$1.80$}	\\
			122	&	318	&	$34.2$	                & 19.4 & 19.9 & 19.5 & 20.3 & &	\\
			\hline
			130	&	338	&	$88.9$	                & 18.3 & 18.8 & 18.5 & 19.1 & \multirow{2}{*}{$1.61$} & \multirow{2}{*}{$1.70$}	\\
			130	&	340	&	$89.1$	                & 18.4 & 18.8 & 18.5 & 19.1 & &	\\
			\hline
			136	&	354	&	$227$	                & 17.5 & 18.0 & 17.7 & 18.2 & \multirow{2}{*}{$1.54$} & \multirow{2}{*}{$1.60$}	\\
			136	&	356	&	$227$	                & 17.6 & 18.1 & 17.7 & 18.2 & &	\\
			\hline\hline
		\end{tabular}
		\caption{Energy level shifts $\Delta\varepsilon_{ns}$ in neutral atoms with $ns$ valence electrons and values of the cut-off parameter $b_{ns}$ in the effective potential Eq. (\ref{Eff_Pot_E1}). For comparison, we present also values of this parameter $b$ for $1s$, $2s$ and $2p_{1/2}$ states of hydrogen-like ions.}\label{E1_valence}
\end{table*}

%Table IX
\begin{table*}[htb]
		\begin{tabular}{cc|c|c|c|c|c|c|c|c}
			\hline\hline
			\multirow{2}{*}{$Z$} & \multirow{2}{*}{$A$} & \multirow{2}{*}{$\beta_2$} & $-\frac{\Delta\tilde\varepsilon_{ns}}{B(E2)c_{ns}^2}$ & $\tilde b_{ns}$ & $\tilde b_{1s}$ & $\tilde b_{2s}$ & $\tilde b_{2p_{1/2}}$ &  \multirow{2}{*}{$\delta_R\tilde{b}_s$} &  \multirow{2}{*}{$\delta_R\tilde{b}_{p_{1/2}}$}  \\
			&&& ($10^{-9}{\rm fm}^{-4}$) & (${\rm fm}$) & (${\rm fm}$) & (${\rm fm}$) & (${\rm fm}$) & & \\
			\hline
			26	&	54	& 0.195 &	$ 0.0214 $	& 120  & 120  & 120  & 132 & \multirow{2}{*}{$13.6$} & \multirow{2}{*}{$17.0$}	    \\
			26	&	56	& 0.239 &	$ 0.0207 $	& 121  & 121  & 121  & 133 & &	    \\
			\hline
			34	&	74	& 0.302 &	$ 0.0448 $	& 114  & 114  & 114  & 124 & \multirow{2}{*}{$12.8$} & \multirow{2}{*}{$15.8$}	    	    \\
			34	&	76	& 0.309 &	$ 0.0442 $	& 114  & 115  & 114  & 125 & &	    \\
			\hline
			44	&	102	& 0.244 &	$ 0.0952 $	& 108  & 109  & 108  & 118 & \multirow{2}{*}{$11.8$} & \multirow{2}{*}{$14.3$}      \\
			44	&	104	& 0.257 &	$ 0.0939 $  & 109  & 109  & 109  & 118 & &	    \\
			\hline
			46	&	108	& 0.243 &	$ 0.109 $   & 107  & 108  & 107  & 117 & \multirow{2}{*}{$10.8$} & \multirow{2}{*}{$13.0$}	    \\
			46	&	110	& 0.257 &	$ 0.108 $	& 108  & 108  & 108  & 117 & &	    \\
			\hline
			62	&	152	& 0.306 &	$ 0.338 $	& 96.9 & 97.6 & 96.9 & 105 & \multirow{2}{*}{$9.68$} & \multirow{2}{*}{$11.5$}	    \\
			62	&	154	& 0.341 &	$ 0.335 $   & 97.2 & 97.9 & 97.2 & 105 & &	    \\
			\hline
			66	&	162	& 0.341 &	$ 0.445 $	& 94.4 & 95.2 & 94.4 & 102 & \multirow{2}{*}{$9.52$} & \multirow{2}{*}{$11.2$}	    \\
			66	&	164	& 0.348 &	$ 0.441 $   & 94.7 & 95.4 & 94.7 & 102 & &      \\
			\hline
			70	&	172	& 0.330 &	$ 0.586 $	& 92.0 & 92.8 & 92.0 & 99.2 & \multirow{2}{*}{$8.94$} & \multirow{2}{*}{$10.5$}	    \\
			70	&	174	& 0.325 &	$ 0.580 $	& 92.2 & 93.0 & 92.2 & 99.5 & &	    \\
			\hline
			74	&	184	& 0.235 &	$ 0.764 $	& 89.8 & 90.7 & 89.8 & 96.7 & \multirow{2}{*}{$8.25$} & \multirow{2}{*}{$9.65$}	    \\
			74	&	186	& 0.224 &	$ 0.757 $	& 90.0 & 90.9 & 90.0 & 97.0 & &	    \\
			\hline
			90	&	228	& 0.230 &	$ 2.36 $	& 80.3 & 81.5 & 80.3 & 86.0 & \multirow{2}{*}{$6.99$} & \multirow{2}{*}{$8.00$}	    \\
			90	&	230	& 0.244 &	$ 2.35 $	& 80.4 & 81.6 & 80.5 & 86.2 & &	    \\
			\hline
			92	&	234	& 0.272 &	$ 2.73 $	& 79.1 & 80.4 & 79.2 & 84.7 & \multirow{2}{*}{$6.90$} & \multirow{2}{*}{$7.90$}	    \\
			92	&	236	& 0.282 &	$ 2.71 $	& 79.3 & 80.5 & 79.4 & 84.9 & &	    \\
			\hline
			98	&	250	& 0.299 &	$ 4.29 $	& 75.7 & 77.0 & 75.8 & 80.9 & \multirow{2}{*}{$6.54$} & \multirow{2}{*}{$7.47$}	    \\
			98	&	252	& 0.304 &	$ 4.26$	    & 75.8 & 77.1 & 75.9 & 81.0 & &	    \\
			\hline
			106	&	234	& 0.272 &	$ 2.73 $	& 79.1 & 80.4 & 79.2 & 84.7 & \multirow{2}{*}{$5.97$} & \multirow{2}{*}{$6.75$}	    \\
			106	&	236	& 0.282 &	$ 2.71 $	& 79.3 & 80.5 & 79.4 & 84.9 & &	    \\
			\hline
			114	&	234	& 0.272 &	$ 2.73 $	& 79.1 & 80.4 & 79.2 & 84.7 & \multirow{2}{*}{$5.58$} & \multirow{2}{*}{$6.24$}	    \\
			114	&	236	& 0.282 &	$ 2.71 $	& 79.3 & 80.5 & 79.4 & 84.9 & &	    \\
			\hline
			122	&	234	& 0.272 &	$ 2.73 $	& 79.1 & 80.4 & 79.2 & 84.7 & \multirow{2}{*}{$5.38$} & \multirow{2}{*}{$5.95$}	    \\
			122	&	236	& 0.282 &	$ 2.71 $	& 79.3 & 80.5 & 79.4 & 84.9 & &	    \\
			\hline
			130	&	234	& 0.272 &	$ 2.73 $	& 79.1 & 80.4 & 79.2 & 84.7 & \multirow{2}{*}{$5.08$} & \multirow{2}{*}{$5.56$}	    \\
			130	&	236	& 0.282 &	$ 2.71 $	& 79.3 & 80.5 & 79.4 & 84.9 & &	    \\
			\hline
			136	&	234	& 0.272 &	$ 2.73 $	& 79.1 & 80.4 & 79.2 & 84.7 & \multirow{2}{*}{$4.65$} & \multirow{2}{*}{$5.04$}	    \\
			136	&	236	& 0.282 &	$ 2.71 $	& 79.3 & 80.5 & 79.4 & 84.9 & &	    \\
			\hline\hline
		\end{tabular}
		\caption{Energy level shifts in neutral atoms calculated with $ns$ valence electron wave function and values of cut-off parameter $b$ which determine the effective potential (\ref{Eff_Pot_E2}).}
		\label{E2_valence}
	\end{table*}
\end{center}

\end{document}